\documentclass[aip,twocolumn,jcp,superscriptaddress]{revtex4}
\usepackage{graphicx}
\usepackage{amssymb}
\usepackage{amsmath}

\usepackage{color}

\newcommand{\R}{\mathbf{r}}
\newcommand{\mycomment}[1]{{#1}}

\def\atau{\tau}
\def\tKS{{\text{KS}}}
\def\tGKS{{\text{GKS}}}
\begin{document}

\title{Laplacian-dependent models of the kinetic energy density: Applications in subsystem density functional theory with meta-generalized gradient approximation functionals}

\author{Szymon \'Smiga}
\affiliation{Istituto Nanoscienze-CNR, Italy}
\affiliation{Institute of Physics, Faculty of Physics, Astronomy and Informatics,
Nicolaus Copernicus University, Grudziadzka 5, 87-100 Torun, Poland}
\affiliation{Center for Biomolecular Nanotechnologies @UNILE, 
Istituto Italiano di Tecnologia (IIT), Via Barsanti, 73010 Arnesano (LE), Italy}
\author{Eduardo Fabiano}
\affiliation{Institute for Microelectronics and Microsystems (CNR-IMM), Via Monteroni, Campus Unisalento, 73100 Lecce, Italy}
\affiliation{Center for Biomolecular Nanotechnologies @UNILE, 
Istituto Italiano di Tecnologia (IIT), Via Barsanti, 73010 Arnesano (LE), Italy}
\author{Lucian A. Constantin}
\affiliation{Center for Biomolecular Nanotechnologies @UNILE, 
Istituto Italiano di Tecnologia (IIT), Via Barsanti, 73010 Arnesano (LE), Italy}
\author{Fabio Della Sala}
\affiliation{Institute for Microelectronics and Microsystems (CNR-IMM), Via Monteroni, Campus Unisalento, 73100 Lecce, Italy}
\affiliation{Center for Biomolecular Nanotechnologies @UNILE, Istituto 
Italiano di Tecnologia (IIT), Via Barsanti, 73010 Arnesano (LE), Italy}

\begin{abstract}
The development of semilocal models for the kinetic energy density (KED)
is an important topic in density functional theory (DFT).
This is especially true for subsystem DFT, where these models are
necessary to construct the required non-additive embedding
contributions. In particular, 
these models can also be efficiently employed to replace the exact KED in meta-Generalized Gradient Approximation (meta-GGA)
 exchange-correlation functionals
allowing to extend the subsystem DFT applicability to the meta-GGA level of theory.

Here, we present a two-dimensional scan of semilocal KED models as linear functionals of 
the reduced gradient and of the reduced Laplacian, for atoms and weakly-bound molecular systems.
We find that several models can perform well but in any case
the Laplacian contribution is extremely important
to model the local features of the KED. Indeed a simple
model constructed as the sum of Thomas-Fermi KED and 1/6 of the 
Laplacian of the density yields the best accuracy
for atoms and weakly-bound molecular systems.
These KED models are tested within subsystem DFT with various meta-GGA exchange-correlation functionals for non-bonded systems, showing a 
good accuracy of the method.    
\end{abstract}

\maketitle

\section{Introduction}\label{sec:int}
Advances in Kohn-Sham density functional theory 
\cite{ks,dftbook} are strictly related to
the availability of accurate and efficient approximations for 
the exchange-correlation (XC) functional. 
Over the years, many XC functionals have been proposed \cite{scus_rev,perdewAIPCP2001}.
The simplest ones, namely the local density approximation (LDA)
\cite{ks} and the generalized gradient approximations
\cite{langrethPRB1983} (GGAs), 
display an explicit dependence on the electron density
(and its gradient). Thus, they are practical tools within
the KS scheme. 
More advanced methods are developed introducing additional
input ingredients, that are not explicit functionals of the density.
The most notable examples in this context are the hybrid
\cite{pbe0_2,arbuznikovJSC2007} and the meta-GGA 
\cite{tpss,kinrev} XC functionals. 
The former include a fraction of Hartree-Fock exchange energy and are
thus true non-local approaches. The meta-GGA functionals instead
have the general form
\begin{equation}\label{eq:1}
E_{xc} = \int e_{xc}(\rho(\R),\nabla\rho(\R), \tau^\tKS(\R))d^3\R\ ,
\end{equation}
where
\begin{equation}\label{eq:2}
\tau^\tKS(\R) = \frac{1}{2}\sum_i^\mathrm{occ.}\left|\nabla\phi_i(\R)\right|^2
\end{equation}
is the positive defined kinetic energy density (KED)
(with $\phi_i$ being the 
occupied KS orbitals). This ingredient, which characterizes the
meta-GGA level of theory, allows the functionals to distinguish different
density regions and to achieve high accuracy for a broad range of systems
and properties \cite{tpss,revtpss,js,tpssloc,bloc_hole,alpha,vsxc,schmider98,m06l,regtpss,mggms_1,mgga_ms2,
m11l,xiaosolid,sun13,stare04,adamo00,riley07,scan,sunNC2016,m15l,tao16,kinrev}. 
For these reasons meta-GGA XC functionals
are gaining increasing popularity \cite{tpss,revtpss,js,tpssloc,alpha,vsxc,m06l,regtpss,mggms_1,mgga_ms2,
m11l,mbeef_2014,scan,m15l,b97mv_2015,tao16,satpss_2016}. 
For a recent Review of meta-GGAs functionals, see Ref. \onlinecite{kinrev}.

The positive defined kinetic energy density is a semilocal quantity
which is much easier to compute and manage than a fully non-local
one such as the exact exchange energy, but it is not an explicit functional
of the electron density. Thus, meta-GGA functionals cannot be
straightforwardly used in the KS scheme
\cite{kummel2008,arbuznikov03,zaharievJCP2013}. 
Instead, they are usually implemented
in the generalized KS (GKS) framework \cite{seidl96}. 
This allows to exploit in the
best way the semilocal nature of the functional and
obtain a favorable computational cost. However, it may pose some difficulties
since, in contrast to the KS one, the GKS XC potential is not local
\cite{kinrev,zaharievJCP2013}.

One relevant case where the non-local nature of the potential is particularly
troublesome is the subsystem formulation of DFT \cite{wesorev,subdft_rev,krishtal15}, 
where the nearsightedness principle \cite{prodan}
is employed to partition a complex systems into simpler and smaller subsystems whose
interaction is described by embedding potentials.
In fact, within subsystem DFT the total ground-state electron density $\rho$
of a given system can be written as
\begin{equation}
\rho(\R) = \sum_{I=1}^M\rho_I(\R)\quad ;\quad \rho_I(\R)=\sum_{i=1}^{N_I}\left|\phi^I_i(\R)\right|^2\ ,
\end{equation}
where $M$ is the number of subsystems, $N_I$ is the number of occupied KS
orbitals in the $I$-th subsystem and $\phi^I_i$ are the KS
orbitals of the $I$-th subsystem, which are obtained by solving the
coupled equations \cite{wesowarh93,Wesolowski199671}
\begin{equation}\label{ksced}
\left[-\frac{1}{2}\nabla^2+v_s^{I}(\R)+v^I_{emb}(\R)\right]\phi^I_i(\R)=\epsilon_i^I\phi^I_i(\R) .
\end{equation}
Here $v_s^{I}$ is the ordinary KS potential relative to subsystem $I$,
whereas 
\begin{equation}\label{e3}
v^I_{emb}= \sum_{J\neq I} v_H^J + \frac{\delta T_s^{nadd}}{\delta\rho_I} + \frac{\delta E_{xc}^{nadd}}{\delta\rho_I}
\end{equation}
is the embedding potential which takes into account the presence of the
other subsystems.
Equation (\ref{e3}) includes the classical Coulomb potential 
$v_H$ (electron-nuclei and electron-electron terms) 
of all the other subsystems.
Moreover, it contains the functional derivatives of the 
non-additive kinetic and XC functionals
\begin{eqnarray}
\label{tnadd_eq}
T_s^{nadd} & = & T_s\left[\sum_{I=1}^M\rho_I\right]-\sum_{I=1}^MT_s\left[\rho_I\right] \\
\label{xcnadd_eq}
E_{xc}^{nadd} & = & E_{xc}\left[\sum_{I=1}^M\rho_I\right]-\sum_{I=1}^ME_{xc}\left[\rho_I\right]\ .
\end{eqnarray}
The non-additive terms of Eqs. (\ref{tnadd_eq}) and (\ref{xcnadd_eq}) 
as well as the related 
functional derivatives cannot be calculated in a direct manner for
functionals whose dependence on the density is not known
explicitly. This is always the case for kinetic energy,
which therefore requires an appropriate semilocal approximation
\cite{gotz09,wesolowski97hyd,wesolowski96fhnch,apbe,apbek,fde_lap,tran02link,weso_chap}.
No problem arises instead for LDA or GGA XC functionals.
To go beyond the GGA level of theory, however, 
the standard subsystem DFT formalism based on the
KS scheme cannot be easily applied.

Recently, several advances of subsystem DFT have been proposed
\mycomment{such as the use of subsystems' fractional occupations \cite{fde_fractional}, the use of orbital-dependent functionals \cite{fde_lhf}, and the embedding of wave-functions in DFT \cite{wesolowski08,Dresselhaus2015}. Moreover, particularly interesting for the present work are}
an extension to the GKS framework \cite{fde_hybrid} and applications of subsystem DFT calculations
using meta-GGA XC functionals \cite{fdemeta,ramos15}. 
In Ref. \onlinecite{fdemeta}, we presented a formal approach for subsystem DFT with meta-GGAs: 
the operational equations are analogous to 
Eq. (\ref{ksced}) but $v_s$ is replaced by the GKS potential 
of the subsystem while the XC embedding contribution
[Eq. (\ref{xcnadd_eq}) and the related derivative]
is approximated by a proper semilocal expression.
This is obtained by substituting, in the
meta-GGA XC functional [Eq. (\ref{eq:1})], the
orbital-dependent KED $\tau^\tKS$
with an approximated semilocal model $\atau$. This computational approach 
has proven to be remarkably accurate, in spite of the
well known difficulty of describing the kinetic energy with
semilocal approximations and the fact that
the XC functional depends non-linearly on the KED. 
The reason may be traced back to the
fact that in non-additive contributions a prominent role is played
by valence regions, whereas the more complicated core contributions
to the kinetic energy are not much relevant.
Nevertheless, in previous studies only one meta-GGA XC functional
was considered [the Tao-Perdew-Staroverov-Scuseria \cite{tpss} (TPSS) one]
together with two simple models for the kinetic energy density.
Thus, a deep understanding of the methodology is still lacking.

In this paper we aim at \mycomment{improving this work.}
Therefore, in Section \ref{sec:meth} we introduce 
a whole family of semilocal KED models and in
Section \ref{sec:results} we test them on different
properties and systems, including subsystem DFT calculations
using different meta-GGA XC functionals.
In this way, we can obtain not only 
a full assessment of the method and of the 
various KED models but also a \mycomment{more complete} understanding of the
basic features that are required to construct successful
semilocal KED models.
\mycomment{Note that we restrict our study to quite simple
KED models, displaying a simple dependence on the gradient and
a linear one on the Laplacian of the density. This choice allows
to obtain reasonably accurate results avoiding, at the same time, 
an excessive complexity in the models that would prevent
the possibility of a detailed analysis and of a quite complete
understanding of the underlying physics. Of course, in search of
very accurate and realistic KED models one is required, in general, 
to go beyond simple functional forms as those considered here.
However, this increases significantly the complexity of the problem
also because the exploration of the corresponding huge
fitting space calls for the use of advanced computational techniques (e.g.
machine learning). Hence, we leave this task for future work.}

\section{Semilocal KED models and methodology}\label{sec:meth}
Most of semilocal models for $\tau^\tKS$
can be written as

\begin{equation}
\label{eq:tauGE2gen}
\atau = \tau^{\text{TF}} \left( F(s) + b\, q \right)\ ,
\end{equation}
where  $\tau^\text{TF}=(3/10)(3\pi^2)^{2/3}\rho^{5/3}$ is the
Thomas-Fermi (TF) KED \cite{thomas26,fermi28,fermi27},
$s=|\nabla\rho|/[2(3\pi^2)^{1/3}\rho^{4/3}]$ is the reduced gradient, 
$q=\nabla^2\rho/[4(3\pi^2)^{2/3}\rho^{5/3}]$ is the reduced Laplacian,
 $F(s)$ is a GGA enhancement factor (i.e. a function of $s$), and $b$ is a coefficient.
Functionals with the general form in Eq. (\ref{eq:tauGE2gen}) are named
Laplacian-Level meta-GGA (LLMGGA) kinetic energy functionals.
If $b=0$, instead, we have simple GGA kinetic energy functionals.
  
Among the KED approximations based on Eq. (\ref{eq:tauGE2gen}), 
we mention some of particular relevance for the present work:
\begin{itemize}
\item [-] The simple TF plus von Weizs\"acker approximation (TFW)
obtained setting $b=0$ and $F(s)=1+F^\text{W}(s)$ with $F^\text{W}=(5/3)s^2$, so that
$\tau^{\text{W}}=\tau^{\text{TF}}F^\text{W}(s) $ is the von 
Weizs\"acker KED \cite{vw}.
The TFW (GGA) model describes correctly the density-tail asymptotic region 
in different applications \cite{QHD16}. However, it does not yield
in general a very accurate KED \cite{fdemeta}.
\item[-]The second-order gradient expansion (GE2) 
\cite{kirzhnitz1967field,gea2}, 
with $F(s)=F^{\text{GE2}}(s)=1+\mu^{\text{GE2}}s^2$,
$\mu^{\text{GE2}}= 5/27$  and $b=20/9$.
The GE2 (LLMGGA) describes correctly the slowly-varying density limit.
\item[-] The Yang's general formula, obtained from the one-body 
Green's function in the mean-path approximation considering  the
Feynman path-integral approach \cite{yang1986gradient}. This has
\begin{equation}
F(s)=F^{\text{Y}}(s)=1+\frac{5-3 b}{9}s^2 \label{eq:yang} \, , 
\end{equation}
where $b$ is the coefficient in Eq. (\ref{eq:tauGE2gen}).
The Yang's LLMGGA model describes accurately the KED of atoms and 
molecules \cite{yang86}, in particular when $b=10/9$, which
corresponds to $F^\text{Y}(s)=F^{\text{GE2}}(s)$.
\item[-] The modified GE2 (MGE2) \cite{mge2} approximation, with 
$F(s)=F^{\text{MGE2}}(s)=1+\mu^{\text{MGE2}} s^2$, 
$\mu^{\text{MGE2}}= 0.2389$, and $b=20/9$.
The MGE2 (LLMGGA) describes accurately large neutral atoms.
\item[-] The APBEK and  revAPBEK GGA functionals \cite{apbe,apbek} with 
$b=0$ and enhancement factor
\begin{equation}\label{fsr}
F^{[rev]APBEK}(s) =1+\left(\frac{1}{\kappa} + \frac{1}{\mu^{\text{MGE2}} s^2} \right)^{-1} \ ,
\end{equation}
where $\kappa=0.804$ for APBEK and $\kappa=1.245$ for revAPBEK.
These functionals, in particular revAPBEK, give very accurate 
non-additive kinetic embedding energies for non-covalently interacting
systems \cite{apbe,apbek}.
\item[-] The $\tau^{L}$ LLMGGA model, having $F(s)=F^{\text{revAPBEk}}(s)$ and
$b=20/9$. This is an extension of the revAPBEK functional, designed to
yield the same kinetic energy but a better description of the KED.
Indeed, it has been shown to perform well, in 
subsystem DFT calculations, when it is used to
replace $\tau^\tKS$ inside a TPSS meta-GGA XC functional
\cite{fdemeta}.
\end{itemize}

Note that, the Laplacian term in Eq. (\ref{eq:tauGE2gen}) gives no 
contribution to the total kinetic energy \mycomment{(for finite systems)}
nor to the kinetic potential\mycomment{; see appendix \ref{appendixa}.} 
Thus, it is often omitted in many applications.
However, in Refs. \onlinecite{yang86,alva07,fdemeta} it has been shown 
that the Laplacian term is a very important quantity
to describe accurately the spatial dependence of the KED. 
Nevertheless, because of the presence of this term, most of the 
models based on Eq. (\ref{eq:tauGE2gen}) violate 
one or both of the following exact conditions:
\begin{eqnarray}
\tau^\tKS &\ge& 0 \label{eq:pos} \\  
\tau^\tKS &\ge& \tau^\text{W} \ . \label{eq:paul}
\end{eqnarray}

In fact, as $q$ can be negative in some regions of space, e.g. near 
the nucleus, then Eq. (\ref{eq:pos}) is easily violated if $b\neq 0$.
Instead, the condition in Eq. (\ref{eq:paul}) is harder to 
satisfy and it is actually respected only by few accurate functionals 
\cite{lucianLL,karaPRB09,kara13,cancioJCP16} and by the simple TFW approximation.
To avoid this drawback, models based on Eq. (\ref{eq:tauGE2gen})
can be {\it regularized} forcing $\atau$ to recover $\tau^\text{W}$ 
near the nucleus \cite{yang86,lucianLL} (see also 
Eqs. (\ref{eq:z1}) and (\ref{eq:z2}) later on).
In this way, however, the Laplacian term will not integrate to zero, thus the regularization
will change the kinetic energy and its functional derivative.

To assess the performance of these
models and especially to understand the role of
the gradient and Laplacian contributions therein, 
in this work we consider the general ansatz
\begin{equation}
\label{eq:tauGE2opt}
\atau(a,b) = \tau^{TF} \left( 1 + a s^2 + b q \right)\ ,
\end{equation}
where $a$ and $b$ are parameters.
Equation (\ref{eq:tauGE2opt}) provides a quite flexible expression
which recovers most of the aforementioned KED models.
On the other hand, Eq. (\ref{eq:tauGE2opt}) does not include 
high-order terms in $s$, such as the one  in Eq. (\ref{fsr}). 
Nevertheless, typical meta-GGA exchange functionals, i.e. TPSS, show a significant 
dependence on the KED only for small values of $s$, see Ref. \cite{tpss}. Thus, the KED needs to be well approximated 
only in slowly-varying density region.
 


We perform a full two-dimensional scan of the parameters 
$a$ and $b$ in Eq. (\ref{eq:tauGE2opt}) in order
to study the possible best performance of the ansatz 
and understand the role of the different contributions
depending on $s^2$ and $q$.
To this purpose, we consider, for each system, the following
error indicators (functions of $a$ and $b$)
\begin{eqnarray}
\delta_z(a,b)      &=& \frac{1}{N} \int   \rho(\R) \left  | z^\tKS     - z(a,b)         \right | d^3\R \ , \label{eq:err1} \\
\delta_\alpha(a,b)  &=& \frac{1}{N} \int  \rho(\R)  \left | \alpha^\tKS - \alpha(a,b)    \right | d^3\R\ ,\label{eq:err2}  \\
\delta_x(a,b)      &=& \frac{1}{N} \int  \rho(\R)  \big  | F_x(s,z^\tKS,\alpha^\tKS) \nonumber \\
                  & & \quad\quad -F_x(s,z(a,b),\alpha(a,b)   )         \big | d^3\R \ , \label{eq:err3}
\end{eqnarray}
where $N$ is the number of electrons of the system, $F_x$
denotes the TPSS exchange enhancement factor \cite{tpss},
and 
\begin{eqnarray}
z^\tKS      &=& {\tau^\text{W}}/{\tau^\tKS} \ , \\
\alpha^\tKS &=& (\tau^\tKS-\tau^\text{W})/{\tau^\mathrm{TF}} \ , \\
z(a,b)      &=& 1-\frac{{\rm max}\left ( 0, \atau(a,b)  - \tau^{\text{W}}  \right)}{\atau(a,b)} \label{eq:z1} \ , \\
\alpha(a,b) &=&  \frac{ {\rm max}\left ( 0, \atau(a,b)  - \tau^{\text{W}}  \right )}{\tau^\text{TF}}\ .  \label{eq:z2}
\end{eqnarray} 
The quantities $z^\tKS$ and $\alpha^\tKS$
are the main $\tau^\tKS$-dependent variables
used in the construction of the meta-GGA exchange  
\cite{kinrev}. 
In Eqs. (\ref{eq:z1}) and  (\ref{eq:z2})  
they are regularized in order to satisfy the constraint of Eqs. 
(\ref{eq:pos}) and (\ref{eq:paul}).
As previously mentioned, the regularization is very important near 
the core if $b\ne0$; see also section Sec. \ref{sec:scan}.

The indicators defined in Eqs. (\ref{eq:err1})-(\ref{eq:err3}) 
express how much an approximated KED impacts on average on 
the accuracy of $z^\tKS$ and $\alpha^\tKS$ as well as on a typical 
exchange enhancement factor.
Thus, they provide a measure on the expectable accuracy of each model
for subsystem DFT calculations.

Finally, to obtain an overall assessment,
we average, for each indicator, over $M$ different systems as
\begin{equation}
\Delta_\theta(a,b)= \frac{1}{M} \sum_i^M \delta_\theta(a,b) \label{eq:errm} \ ,
\end{equation}
where $\theta=z,\alpha,$ or $x$.

\section{Computational details}
\label{sec:comp}
\subsection{Two-dimensional scan}
To study the model of Eq. (\ref{eq:tauGE2opt}), we have
performed a two dimensional scan over the parameters space
$(a,b)\in[0,2]\times[0,4]$, considering the following training set: 
Ne$_2$, Ne-Ar,(CH$_4$)$_2$ (WI);
(H$_2$S)$_2$, CH$_3$Cl-HCl, CH$_3$SH-HCN (DI); (HF)$_2$, (H$_2$O)$_2$, HF-HCN 
(HB); AlH-HCl,LiH-HCl, BeH$_2$-HF (DHB); NH$_3$-F$_2$, C$_2$H$_2$-ClF, 
H$_2$O-ClF (CT).
For this set we have calculated the error indicators
$\delta_z$, $\delta_\alpha$, and $\delta_x$ defined by 
Eqs. (\ref{eq:err1})-(\ref{eq:err3}).
All quantities have been calculated on densities obtained from standard 
KS-DFT supermolecular calculations, using the PBE \cite{pbe} XC functional.
\mycomment{In some cases, aprroximate KED have been regularized
imposing the constraint of Eq. (\ref{eq:paul}), i.e. 
$\tau^{approx(reg.)} = \max(\tau^{approx},\tau^{\text{W}})$; see Eqs.
(\ref{eq:z1}), (\ref{eq:z2}), and (\ref{tau3}).}

All molecular calculations have been performed using a locally modified
version of the TURBOMOLE \cite{turbomole,turbo_review} program package. 
In order to guarantee a good accuracy of the results and to 
minimize numerical errors, the
supermolecular def2-TZVPPD \cite{def2tzvpp,furchepol}
basis set was employed in all calculations, together with very
accurate integration grids (\texttt{grid 7}, \texttt{radsize 14}).
For all KS-DFT calculations tight convergence criteria were enforced,
corresponding to maximum deviations in density matrix elements of 
$10^{-7}$ a.u. 

For the noble atoms we used a fully numerical code \cite{QHD16} and the LDA functional.

\subsection{Subsystem DFT}
To assess the performance of the various $\atau$ models,
we have employed them to carry on subsystem DFT calculations
with meta-GGA XC functionals. In this case, we have 
computed the non-additive XC meta-GGA terms using the
TPSS \cite{tpss}, revTPSS \cite{revtpss},
BLOC \cite{bloc,tpssloc,bloc_hole}, meta-VT\{8,4\} \cite{delCPL12},
and MGGA\_MS2 \cite{mgga_ms2} functionals.
Note that the latter functional uses a GGA correlation
expression, so the approximation concerns in this case only the exchange term.
For all subsystem DFT calculations, we have considered a partition in
two subsystems and we have performed a full relaxation of embedded 
ground-state electron densities 
through freeze-and-thaw cycles \cite{wesorev,Wesolowski199671}.
These calculations have been performed using the FDE script \cite{fde_hybrid} of the TURBOMOLE program \cite{turbomole}
considering as convergence criterion 
the difference of the dipole moments of the embedded subsystems
($<10^{-3}$ a.u.). 
To compute the non-additive kinetic contributions [Eq. (\ref{tnadd_eq})]
the revAPBEK kinetic functional \cite{apbe,apbek} has been employed.
As to the non-additive meta-GGA XC term, we have employed the
computational procedure described in Ref. \onlinecite{fdemeta}
(i.e. substitution of $\tau^\tKS$ with a semilocal model)
using the $\atau$ models described in Section \ref{sec:meth}.
Such calculations have been performed on 
the following sets of non-covalent complexes:
\begin{itemize}
 \item[{\bf WI}:] weak interaction (He-Ne, He-Ar, Ne$_2$, Ne-Ar, CH$_4$-Ne,
C$_6$H$_6$-Ne, (CH$_4$)$_2$)
 \item[{\bf DI}:] dipole-dipole interaction  ((H$_2$S)$_2$, (HCl)$_2$, H$_2$S-HCl, CH$_3$Cl-HCl,CH$_3$SH-HCN, CH$_3$SH-HCl)
 \item[{\bf HB}:] hydrogen bond ((NH$_3$)$_2$, (HF)$_2$, (H$_2$O)$_2$, HF-HCN,
(HCONH$_2$)$_2$, (HCOOH)$_2$)
 \item[{\bf DHB}:] double hydrogen bond (AlH-HCl, AlH-HF, LiH-HCl, LiH-HF,
MgH$_2$-HCl, MgH$_2$-HF, BeH$_2$-HCl, BeH$_2$-HF)
 \item[{\bf CT}:] charge transfer (NF$_3$-HCN,C$_2$H$_4$-F$_2$,NF$_3$-HCN,
C$_2$H$_4$-Cl$_2$, NH$_3$-F$_2$, NH$_3$-ClF, NF$_3$-HF, C$_2$H$_2$-ClF,
HCN-ClF, NH$_3$-Cl$_2$, H$_2$O-ClF, NH$_3$-ClF).
\end{itemize}
The geometries as well as the reference binding energies were
taken from Refs. \onlinecite{truhlar05a,truhlar05nb,wesolowski96fhnch,fde_ct,dihydrogen}.

The accuracy of the subsystem DFT calculations has been measured considering
the following two quantities.
The first one is the error on the total embedding
energy $\Delta E$, defined as 
\begin{equation}\label{dE}
\Delta E = E_{A+B}[\tilde{\rho}_A,\tilde{\rho}_B]-E^\tGKS[\rho^\tGKS]\ ,
\end{equation}
where $E_{A+B}[\tilde{\rho}_A,\tilde{\rho}_B]$ is the 
total energy obtained from the subsystem DFT calculation,
with  $\tilde{\rho}_A$ and $\tilde{\rho}_B$ being the embedded 
subsystem densities and $E^\tGKS$ being the conventional
GKS total energy of the complex, computed at the
``true'' ground-state density $\rho^\tGKS$ 
\cite{fde_hybrid,apbek,fde_hyb_ene}. 
The second one is the embedding density error defined as
\begin{equation}\label{xi}
\xi=\frac{1000}{N}\int\left|
\Delta \rho(\R) \right|\,d \mathbf{r} \ ,
\end{equation}
where $N$ is the total number of electrons  and $\Delta \rho(\mathbf{r})$ is 
the deformation density
\begin{equation}
\Delta \rho(\mathbf{r}) = \tilde{\rho}_A(\mathbf{r}) + \tilde{\rho}_B(\mathbf{r}) - \rho^\tGKS(\mathbf{r})\ .
\end{equation}
Note that only valence electron densities have been considered in the calculation of density errors \cite{fdemeta}.

Finally, to have an overall indication of the performance of the 
different $\atau$ approximations, we computed, 
within each group of molecules, the the mean absolute
error (MAE) and the mean absolute relative error (MARE),
the latter being referred to reference binding energies \cite{fde_hyb_ene}. 
Furthermore, we considered the two global quantities \cite{apbek}
\begin{eqnarray}
\label{rwmae_eq}
\text{rwMAE}&=&\frac{1}{5} \sum_{i} \left( \frac{MAE_i}{<MAE_i>}\right) \\
\label{rwmare_eq}
\text{rwMARE}&=&\frac{1}{5} \sum_{i} \left( \frac{MARE_i}{<MARE_i>}\right)
\end{eqnarray}
where the sums extend over WI, DI, HB, DHB, CT, and  $<MAE_i>$  
($<MARE_i>$)  is the average MAE (MARE) among
the different methods considered for the class of systems $i$.

\subsection{Embedding energy error decomposition}
\label{appa}
The error on embedding energy can be written \cite{fde_hyb_ene,fdemeta}
\begin{equation}
\Delta E = \Delta D + \Delta T_S + \Delta E_{xc}\ .
\end{equation}
The error components are
\begin{eqnarray}
\Delta D & = & E_{H}\left[\sum_I\rho_I\right] + \widetilde{E}_{xc}\left[\sum_I\rho_I\right] + \\
\nonumber
&&\quad - \left\{E_{H}\left[\rho^\tGKS\right] + \widetilde{E}_{xc}\left[\sum_I\rho^\tGKS\right]\right\} \ ,\\
\Delta T_S & = & \widetilde{T}_s^{nadd}  - T_S[\Phi^\tGKS] -\sum_IT_s[\Phi^I] \ ,\\
\nonumber
\Delta E_{xc} & = & \widetilde{E}_{xc}[\rho^\tGKS] - \sum_I\widetilde{E}_{xc}[\rho_I]  \nonumber \\
              &- & \left\{E_{xc}[\Phi^\tGKS] - \sum_IE_{xc}[\Phi^I]\right\}\ ,
\end{eqnarray}
where $E_H$ is the Hartree energy 
(electron-electron and electron-nuclei classical Coulomb interaction energy),
$\rho^\tGKS$ and $\Phi^\tGKS$ are respectively the density 
and the Slater determinant of the supermolecular system as obtained from a
standard GKS calculation, and the tilde ($\ \widetilde{}\;$) denotes
that an approximate functional form is considered.

%

\section{Results}\label{sec:results}

\subsection{Two dimensional scan of the $\atau$ model}
\label{sec:scan}
\begin{figure}
\begin{center}
\includegraphics[width=\columnwidth]{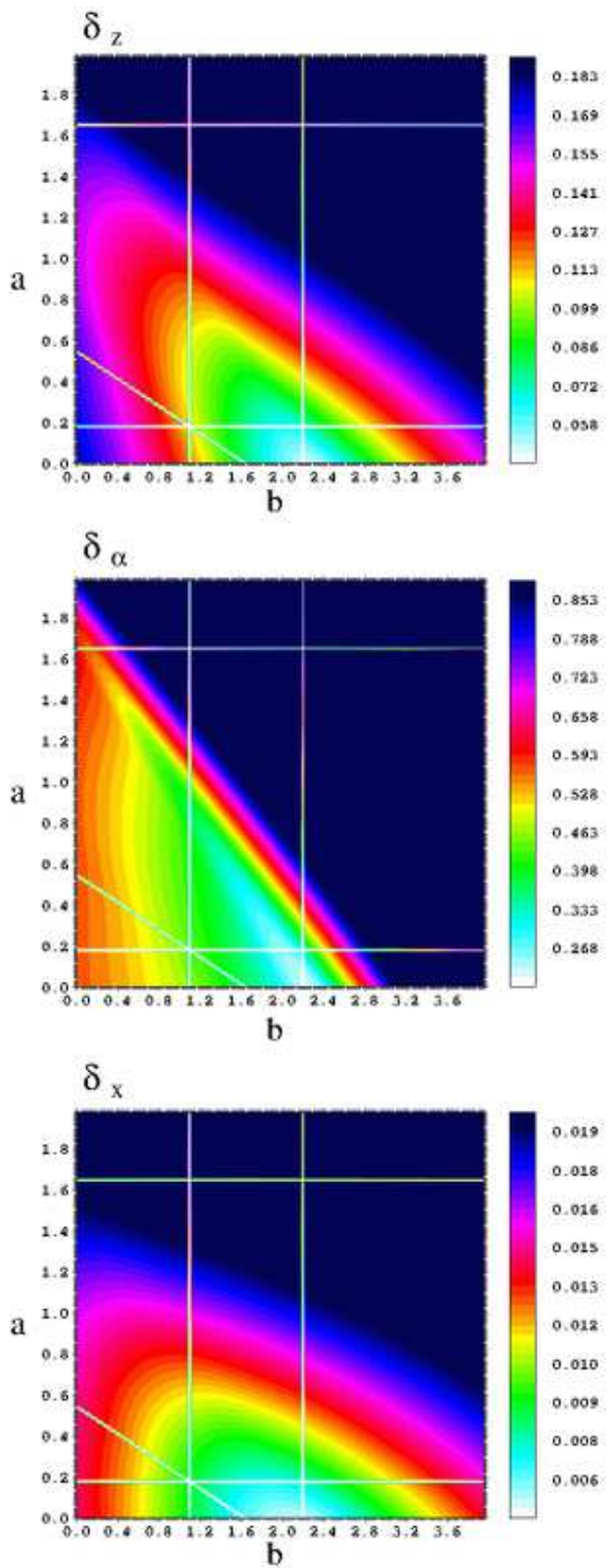}
\caption{\label{figneon} Values of $\delta_z(a,b)$ , $\delta_\alpha(a,b)$  
and $\delta_x(a,b)$ [see Eqs. (\ref{eq:err1})-(\ref{eq:err3})] as functions of the values of the $a$ and $b$ parameters 
in Eq. (\ref{eq:tauGE2opt}) for the Neon atom. 
The white horizontal lines denote $a=5/27$ and $a=5/3$. 
The white vertical lines denote $b=10/9$ and $b=20/9$. 
The white diagonal line represents the Yang's formula [Eq. (\ref{eq:yang})]. 
The color map is linear and, in each panel, it is limited to represent values not larger than four times the minimum value. }
 \end{center}
\end{figure}

Figure \ref{figneon} reports the values of the error
indicators $\delta_z$, $\delta_\alpha$, 
and $\delta_x$ [see Eqs. (\ref{eq:err1})-(\ref{eq:err3})]
as computed for the Neon atom in the case of the KED model
of Eq. (\ref{eq:tauGE2opt}). Similar results are found
for Argon and Radon, see Fig. S1 of Ref. \cite{suppmat}.

The most important result of Fig. \ref{figneon} is that the minima for all the three indicators
appear at $a=0$ and $b\approx2.2$: the latter almost coincide with the ``non-empirical'' value of 20/9 (from 
the second-order gradient expansion).
Similar findings have been obtained in Ref. \cite{alva07}, where the value of $b$ had been fitted for different 
GGA KED, in order to reproduce the exact KED for the first 10 atoms (H-Ne) of the periodic table.
The result in Fig. \ref{figneon}, is more general, because we demonstrate that $(a,b)=(0,2.2)$ is the global minimum.

Thus we introduce a new KED model, named regularized  Thomas-Fermi plus Laplacian (TFL),
that is defined by the formulas
\begin{eqnarray}
\tau^{TFL(reg.)} &=& \max \left( \tau^{TFL} , \tau^\text{W} \right) \ , \label{tau3} \\
\tau^{TFL}       &=& \tau^{TF} \left( 1 + \frac{20}{9} q\right)  \label{tau3n} \ .
\end{eqnarray}
We note again that the TFL model can be considered a ``non-empirical'' functional because it is defined with 
the coefficient $b=20/9$ which comes from the second-order gradient expansion
\mycomment{(however, it is not ab initio, because its general form is
  suggested by the numerical analysis of the error indicators)}.
As the TFL model in Eq. (\ref{tau3n}) depends only on the Laplacian and not on the gradient, and 
it is not a LLMGGA kinetic energy functional. The TFL functional, instead, belongs to a different class of functionals 
which can be named Laplacian-Level density approximation (LLDA).
On the other hand, its regularized version, i.e., the TFL(reg.) 
functional in Eq. (\ref{tau3}), shall be considered as a conventional LLMGGA functional, as it also depends on the gradient
of the density through $\tau^\text{W}$.  However, the regularization is only active near 
the core (see hereafter), which is not relevant for molecular
applications. Thus, for simplicity, we will consider both TFL and TFL(reg.) as LLDA functionals, to distinguish them from other
 LLMGGA functionals which depend on the gradient also in the valence region.


Fig. \ref{figneon} shows that the shapes of 
$\delta_z$ and $\delta_x$ are rather similar.
This similarity is indeed not surprising because, 
as mentioned before, the $z$ ingredient, tested in $\delta_z$, is the main 
variable used to construct the meta-GGA TPSS exchange enhancement factor,
which is used in the construction of $\delta_x$.
On the other hand, the $\delta_\alpha$ plot looks quite different. 
Nevertheless, the minimum is still at $(a,b)\approx(0,20/9)$.

\begin{figure}
\begin{center}
\includegraphics[width=\columnwidth]{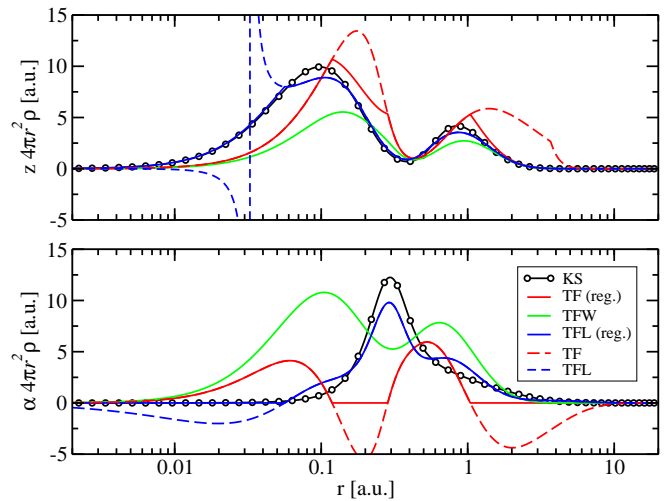}
\caption{\label{figt} Integrand of Eq.  (\ref{eq:err1}), upper panel,  
and Eq.  (\ref{eq:err2}), lower panel, versus the radial distance for the Neon atom for different KED models.}
\end{center}
\end{figure}

Another interesting result of Fig.  \ref{figneon}, is that TF, i.e.  $(a,b)=(0,0)$,  and 
TFW, i.e.  $(a,b)=(5/3,0)$,  yield almost the same
values of $\delta_z$ and  $\delta_\alpha$.
This is due to error cancellation effects, as explained in  
Fig. \ref{figt}, where the integrands of Eq.  (\ref{eq:err1}) 
and Eq.  (\ref{eq:err2}) are reported in the upper and lower
panel, respectively.

Concerning the integrand of $\delta_z$ (upper panel), we see that 
 TFW always underestimates the exact value, whereas TF(reg.) overestimates it, yielding similar global error 
$\delta_z$.
Concerning the integrand of $\delta_\alpha$ (lower panel), 
one can see that TF(reg.) is better than TFW in the range to $r <$ 0.2 a.u 
 and  0.5 a.u. $< r <$ 1.2 a.u., whereas  TFW is better than TF(reg.) in the range 
  0.2 a.u. $< r <$ 0.5 a.u.  and for $r >$ 1.2 a.u, again yielding similar global 
error $\delta_\alpha$.
Clearly, without regularization the results for TF will be much worse (see red dashed-lines in
 Fig.  \ref{figt}).

Figure  \ref{figt} also shows the high accuracy of the TFL(reg.) model which matches the exact KS results
everywhere in the space.
Note that the regularization is very important when $z^\tKS$ needs to be accurately reproduced
near the core region, otherwise, as shown in the upper panel of  Fig.  \ref{figt}, a divergence
(i.e. at $r\approx 0.02$) will appear, as $q$ becomes negative near the core.
On the other hand, regularization is less critical for  $\alpha^\tKS$, as one can see 
comparing TFL and TFL(reg.) in the lower panel Fig.  \ref{figt}.



\begin{figure}
\begin{center}
\includegraphics[width=\columnwidth]{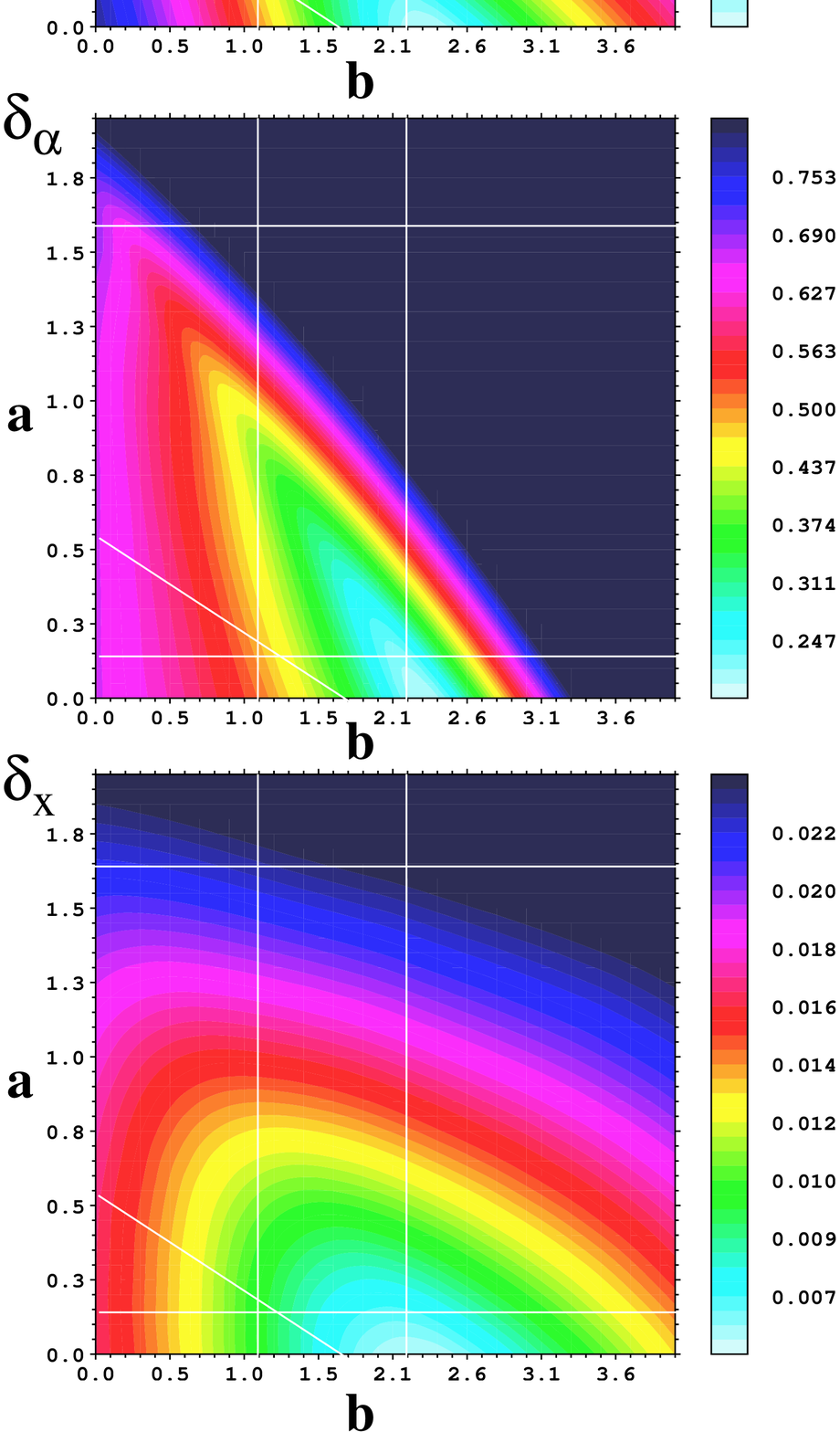}
\caption{\label{fig1} Values  of $\Delta_z(a,b)$ , $\Delta_\alpha(a,b)$  and $\Delta_x(a,b)$ [see Eqs. \ref{eq:errm})] 
as functions of the values of the $a$ and $b$ parameters
 in Eq. (\ref{eq:tauGE2opt}) for a set of molecules (see Sect. \ref{sec:comp}). 
Other details are the same as in Fig. \ref{figneon}.}
\end{center}
\end{figure}

The above results are not limited to atoms, but hold also for molecules.
Figure \ref{fig1} reports a similar investigation as in 
Fig. \ref{figneon} but averaged over a set of molecules (see Section \ref{sec:comp}).
Interestingly, the results for the three indicators are 
all similar to the ones shown in Fig. \ref{figneon}, with
the minima again located at $(a,b)\approx(0,20/9)$.
This finding is quite important and confirms previous results for atoms.

Thus the TFL(reg.) model is the one that yields the best 
(average) performance among the models of the family defined 
by Eq. (\ref{eq:tauGE2opt}).

\subsection{Comparison with other kinetic energy functionals}\label{eq:compkin}

In Table \ref{tab1} the performances of KED from different kinetic energy functionals are compared to TFL: 
we considered LDA, GGA and LLMGGA functionals.
Note that all functionals are regularized, but TFW, mGGA and mGGArev which implicitly satisfy Eq. (\ref{eq:paul}).
\begin{table*}
\caption{\label{tab1} Values of $\Delta_z(a,b)$ , $\Delta_\alpha(a,b)$ and $\Delta_x(a,b)$ 
[see Eq.(\ref{eq:errm})] for different KED models evaluated on a molecular test set (see Section \ref{sec:comp} for details). 
If applicable, also the values of the $a$ and $b$ parameters (as defined in Eq. (\ref{eq:tauGE2opt})) 
and the literature reference are reported for each functional. 
The smallest values of the indicators for each class of functionals are highlighted in bold style.}
\begin{ruledtabular}
\begin{tabular}{lrrrrrrr}
Model & Reg. & $\Delta_z$ & $\Delta_\alpha$ & $\Delta_x$  & a & b & Ref. \\
\hline
\multicolumn{7}{c}{Local Density Approximation (LDA)} \\
TF & Yes & 0.193 & 0.662 &  0.017 & 0 & 0   & [\onlinecite{fermi27}] \\
\multicolumn{7}{c}{Generalized Gradient Approximation (GGA)} \\
TFW & Impl. & \textbf{0.186} &	0.690 &	0.021 &  5/3 & 0 & [\onlinecite{fdemeta}] \\
$\tau^{\text{TF}}(1+\mu_{\text{GE2}}s^2)$  & Yes 
 & 0.189 & \textbf{0.662} & \textbf{0.016} & 5/27 & 0 & [\onlinecite{gea2}] \\
$\tau^{\text{TF}}(1+\mu_{\text{MGE2}}s^2)$ & Yes 
&  0.188 & \textbf{0.662} & 0.017 & 0.2389 & 0 &  [\onlinecite{mge2}] \\
revAPBEK & Yes & 0.188 & \textbf{0.662} & 0.017 & - & 0 &  [\onlinecite{apbek}] \\
TW02 & Yes & 0.189 & 0.663 & 0.017 & - & 0 &  [\onlinecite{TW02}]\\
LC94 & Yes & 0.189 & 0.663 &  0.017 & - & 0 &   [\onlinecite{lc94}]\\
\multicolumn{7}{c}{Laplacian-Level Meta-GGA (LLMGGA)} \\
GE2  & Yes & 0.068 & 0.281  & 0.007 & 5/27   & 20/9 & [\onlinecite{gea2}] \\
MGE2 & Yes & 0.076 & 0.326      & 0.008  & 0.2389 & 20/9 & [\onlinecite{mge2}] \\
$\tau^L$
& Yes &  0.070 &0.241 & 0.008 & - & 20/9 &  [\onlinecite{fdemeta}] \\
$\tau^{\text{TW02}} + \frac{20}{9} \tau^\mathrm{TF}q$ & Yes  & 0.068 & {\bf 0.233} & 0.007  & - & 20/9 &  [\onlinecite{TW02}] \\
$\tau^{\text{LC94}} + \frac{20}{9} \tau^\mathrm{TF}q$ & Yes & 0.068 & 0.236 & 0.007 & - & 20/9 & [\onlinecite{lc94}] \\
mGGA & Impl. & {\bf 0.056} & 0.320 & {\bf 0.006}  & - & -& [\onlinecite{lucianLL}] \\
L0.4 & Yes & 0.189&  0.663 & 0.017 & - & - & [\onlinecite{fde_lap}]\\
L0.6 & Yes & 0.189& 0.663 & 0.017 & - & -  &  [\onlinecite{fde_lap}]\\
mGGArev & Impl. & 0.096 & 0.323 &  0.010 & - & -& [\onlinecite{cancioJCP16}]\ \\
\multicolumn{7}{c}{Laplacian-Level Density Approximation (LLDA)} \\
TFL  & Yes & \textbf{0.050} & \textbf{0.203} & \textbf{0.005} &  0 & 20/9 & this work \\
\end{tabular}
\end{ruledtabular}
\end{table*}
The results in Tab. \ref{tab1} show that TF and all GGAs 
give almost the same results for all indicators, as already discussed in section \ref{sec:scan}. 

Large improvements are only obtained when the Laplacian term is considered.
The mGGA functional of Ref. \cite{lucianLL} yields the best results 
for $\Delta_z$ and $\Delta_x$, while TW02 \cite{TW02} regularized with $(20/9) q$, gives the best results for $\Delta_\alpha$,

The last line of Tab. \ref{tab1} shows that TFL model yields the best results overall, confirming the results of Sect. \ref{sec:scan}.  

\subsection{Subsystem DFT calculations}
In this section we apply the TFW,
$\tau^L$, and $\tau^{TFL}$ [Eq. (\ref{tau3n})]
KED models in subsystem DFT calculations using several
different XC meta-GGA functionals.
This can provide a practical assessment of the
reliability of the different models.
We consider TFW as a simple GGA satisfying Eq. (\ref{eq:paul}), $\tau^L$ as a simple meta-GGA, 
previously considered in Ref. \cite{fdemeta}, and
 $\tau^{TFL}$ as the best functional, as discussed in the previous sections.
 Note that, in subsystem DFT, $\tau^L$ and $\tau^{TFL}$ can be also used {\it without regularization}. In fact, as
discussed in Ref. \cite{fdemeta}, core contributions cancel out in the non-additive quantities. 

\begin{figure}
\begin{center}
\includegraphics[width=0.9\columnwidth]{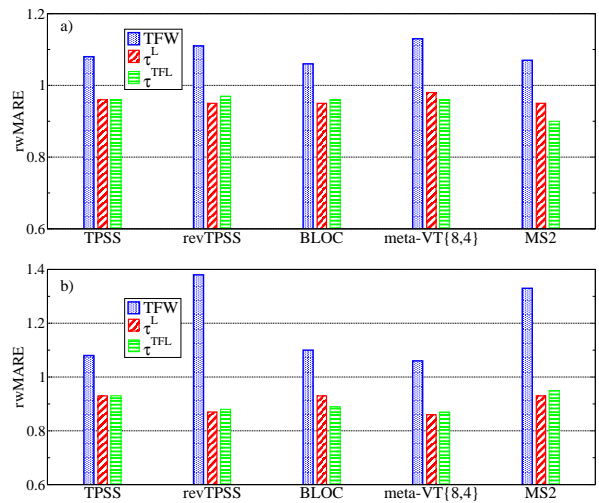}
\caption{\label{figk} Top: the rwMAE [Eq. (\ref{rwmae_eq})] for the embedding
 {\it densities} errors calculated for all the investigated meta-GGA XC functionals and $\atau$ models. 
Bottom: the rwMARE [Eq. (\ref{rwmare_eq})] for the embedding {\it energy} errors calculated for all 
the investigated meta-GGA XC functionals and $\atau$ models.
}
\end{center}
\end{figure}
In Fig. \ref{figk} we report (panel a) the rwMAE [Eq. (\ref{rwmae_eq})] 
for the  embedding {\it densities} errors and  (panel b) the rwMARE 
[Eq. (\ref{rwmare_eq})] for the embedding {\it energy} errors  
evaluated for all the investigated meta-GGA XC functionals and KED models. 
Results for all systems are reported in Ref. \onlinecite{suppmat}.
Concerning the density error, we see that the $\atau$ models containing 
the Laplacian term (i.e. $\tau^L$ and $\tau^{TFL}$)
perform very similarly and accurately, yielding rwMAE
values in the range $0.90-0.98$. 
The $\tau^{TFL}$ is better than  $\tau^L$ only for the MS2 functional, while the performances are comparable for other meta-GGAs.
On the other hand, the $\tau^{TFW}$ model displays a worse
performance, providing rwMAE values oscillating between  $1.06-1.13$.
 
For the embedding energy error, the trend is the same as for the 
density errors: $\tau^L$ and $\tau^{TFL}$ provide the best performance
with quite similar results, but for the BLOC functional where $\tau^{TFL}$ is distinctively better than  $\tau^L$.
In all the cases the TFW model gives definitely poorer results
yielding rwMARE values above 1.1 for all functionals
(to be compared to values of about 0.9 for the $\tau^{TFL}$
and $\tau^L$ models) and it is very inaccurate for revTPSS and MS2. 

For a deeper understanding of these results, it is possible to 
perform an embedding energy error decomposition analysis 
\cite{fde_hyb_ene,fdemeta}
(see subsection \ref{appa} for details). 
This allows to separate the relaxation error $\Delta D$, due to the
fact that the embedding density differs from the reference
Kohn-Sham supermolecular one, from the errors arising from 
the approximations used in the non-additive energy functionals.
The latter can be further divided into a kinetic energy error 
$\Delta T_S$ and an XC error $\Delta E_{xc}$.
The former originates from approximations used in Eq. (\ref{tnadd_eq})
and, in the present calculations, it is the same for all cases 
(we used the revAPBEK kinetic functional in all calculations).
The latter traces back to the use of a semilocal KED model
to compute Eq. (\ref{xcnadd_eq}), for different meta-GGA XC functionals.
Thus, this is the most interesting quantity in the present
context. 

\begin{figure}
\begin{center}
\includegraphics[width=0.9\columnwidth]{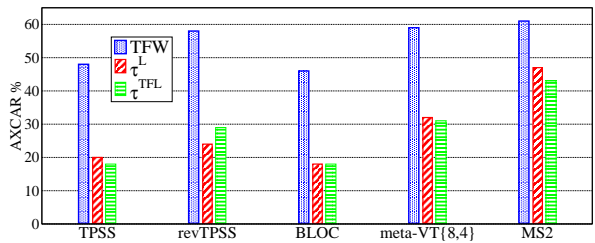}
\caption{\label{fig5}  The AXCAR error [Eq. (\ref{axcar_eq})] calculated for
different methods and $\atau$ models.}
\end{center}
\end{figure}

Hence, we consider in Fig. \ref{fig5} the
average XC absolute ratio (AXCAR)  \cite{fdemeta} defined as 
\begin{equation}\label{axcar_eq}
\text{AXCAR} = \frac{1}{M} \sum^M_{i=1} \frac{|\Delta E_{xc}|}{|\Delta E_{xc}| + |\Delta T_s + \Delta D|} \ ,
\end{equation}
where $M$ is the number of systems in the test set 
(results for all systems are reported in supporting information).
This provides an average absolute (i.e. without error compensation) measure
of the XC contribution to the embedding energy error.
Inspection of the figure shows
that in all cases $\tau^L$ and $\tau^{TFL}$   yield
 AXCAR values much smaller than TFW (AXCAR alwayes larger than 40\%).
Moreover, $\tau^{TFL}$ gives smaller AXCAR than  $\tau^L$, for all functionals, but revTPSS.
These results indicate that the accuracy observed for the 
embedding energy error is not essentially due to some error compensation
effect, but really traces back to a superior
quality of the $\tau^{TFL}$ model in approximating
the kinetic energy density contributions in the
XC meta-GGA non-additive term.
%

\section{Summary and conclusions}\label{sec:con}
In subsystem DFT the non-additive XC term must be an explicit functional
of the density. Therefore, in order to be able to perform
subsystem DFT calculations employing meta-GGA XC functionals,
it is necessary to replace the positive-defined Kohn-Sham KED 
($\tau^\tKS$) with an approximated semilocal KED 
model ($\atau$).
The development of such a model is not as a hard task as the 
general development of accurate kinetic energy functionals because,
in the context of subsystem DFT, only non-additive contributions are important.
Thus, core contributions, which incorporate most of the Pauli kinetic
energy, are not much relevant.
On the other hand, in the present context, there is the need to model
all the spatial features of the KED, since this is employed non-linearly
inside meta-GGA XC functionals and not only to produce a kinetic energy
(via integration) and/or a kinetic potential (via functional derivation).
For this reason, for example, the accurate determination of the proper 
gauge for the KED is a crucial issue in this field.
The development of accurate KED models is therefore a challenging and
interesting topic that has practical relevance in subsystem DFT but
is also of fundamental importance to understand the basic behavior of the
kinetic energy, which is a key quantity in all areas of DFT.

In this work, we studied the problem of developing an accurate semilocal
KED model by considering a flexible ansatz [Eq. (\ref{eq:tauGE2opt})]
depending linearly on the reduced gradient and Laplacian of the
density.
In this way, we found that a Laplacian contribution of the type
$(1/6)\nabla^2\rho$ is essential to fix the correct gauge in the 
KED and thus to provide an accurate description of most of
the features of the exact KDE.
In fact, already a simple model (i.e. the TFL one of Eq. (\ref{tau3}))
built from the Thomas-Fermi KED and the Laplacian correction
(i.e. without any gradient correction), performs remarkably well
for atoms and  molecular complexes.

A thorough assessment work showed that the simple TFL model,
is indeed able
to capture most of the important features of the KED.
Hence, in this context it outperforms all the known gradient 
expansion KED models (e.g. 
the conventional \cite{kirzhnitz1967field}, modified \cite{mge2}, 
path-integral-derived \cite{yang1986gradient} second-order gradient expansions)
as well as many GGA and Laplacian-dependent meta-GGA models.

These findings are important for different topics in DFT.
Clearly they are relevant for subsystem DFT calculations 
with meta-GGA XC functionals.
Thus, they can allow to extend the applicability and accuracy of 
subsystem DFT, permitting to benefit from the advantages of meta-GGA XC
functionals.
Nevertheless, the general results of this work, concerning the
role of different density contributions to the KED and especially
that of the Laplacian, can have a larger impact on future work.
They are in fact extremely relevant for all those studies and applications
where a semilocal model of the KED is important. Here we suggest, for example,
the evaluation of local indicators such as the electron localization 
function (ELF) \cite{elf1,elf2} or the entanglement length \cite{pittalis15}.
These are density indicators but they are actually obtained as
orbital-dependent expressions due to the presence of $\tau^\tKS$.
The use of an appropriate semilocal KED model can turn them into 
true density indicators and largely extend their application domain 
providing as well a deeper understanding of the underlying physics and 
chemistry.

\section*{Acknowledgements}
This work was partially supported by the National Science Center under Grant No.
DEC-2013/11/B/ST4/00771.

\appendix

\section{Total energy and potential of the model KED of Eq. (\ref{eq:tauGE2gen})}
\label{appendixa}
Consider the KED of Eq. (\ref{eq:tauGE2gen}). Note that
it can be written
\begin{equation}
\tau = \tau^\mathrm{TF}F(s) + \frac{3b}{10}\nabla^2\rho = \tau^{GGA} + \frac{3b}{10}\nabla^2\rho\ .
\end{equation}

The corresponding total kinetic energy is
\begin{equation}
T_s = \int \tau^{GGA}d\R + \frac{3b}{10}\int\nabla^2\rho d\R = T_s^{GGA} + \frac{3b}{10}\int\nabla^2\rho d\R\ .
\end{equation}
The last integral can be evaluated via the second Green's identity. Hence,
\begin{equation}\label{a2}
\int\nabla^2\rho d\R =  \int\nabla_\perp\rho dS + \int\rho\nabla^2 1d\R = \int\nabla_\perp\rho dS\ , 
\end{equation}
where $\nabla_\perp\rho$ is the perpendicular component of the
gradient with respect to the surface area element $dS$.
For any finite system, with an exponentially decaying density, the
integral in Eq. (\ref{a2}) vanishes. Therefore, the Laplacian tem does
not contribute to the total kinetic energy.

Concerning the potential we have \cite{kinrev}
\begin{equation}
v_T = \frac{\partial\tau}{\partial\rho} -\nabla\cdot\frac{\partial\tau}{\partial\nabla\rho} + \nabla^2\frac{\partial\tau}{\partial\nabla^2\rho}\ .
\end{equation}
For $\tau$ given by Eq. (\ref{eq:tauGE2gen}), that is
\begin{eqnarray}
v_T & = & \frac{\partial\tau^{GGA}}{\partial\rho}-\nabla\cdot\frac{\partial\tau^{GGA}}{\partial\nabla\rho} + \nabla^2\frac{\partial(3b\nabla^2\rho/10)}{\partial\nabla^2\rho} = \\
\nonumber
& = & v_T^{GGA} + \frac{3b}{10} \nabla^2\frac{\partial\nabla^2\rho}{\partial\nabla^2\rho} = v_T^{GGA} + \frac{3b}{10} \nabla^2 1 = v_T^{GGA}\ .
\end{eqnarray}

 \bibliography{taumeta.bib}

\begin{thebibliography}{91}
\expandafter\ifx\csname natexlab\endcsname\relax\def\natexlab#1{#1}\fi
\expandafter\ifx\csname bibnamefont\endcsname\relax
  \def\bibnamefont#1{#1}\fi
\expandafter\ifx\csname bibfnamefont\endcsname\relax
  \def\bibfnamefont#1{#1}\fi
\expandafter\ifx\csname citenamefont\endcsname\relax
  \def\citenamefont#1{#1}\fi
\expandafter\ifx\csname url\endcsname\relax
  \def\url#1{\texttt{#1}}\fi
\expandafter\ifx\csname urlprefix\endcsname\relax\def\urlprefix{URL }\fi
\providecommand{\bibinfo}[2]{#2}
\providecommand{\eprint}[2][]{\url{#2}}

\bibitem[{\citenamefont{Kohn and Sham}(1965)}]{ks}
\bibinfo{author}{\bibfnamefont{W.}~\bibnamefont{Kohn}} \bibnamefont{and}
  \bibinfo{author}{\bibfnamefont{L.~J.} \bibnamefont{Sham}},
  \bibinfo{journal}{Phys. Rev.} \textbf{\bibinfo{volume}{140}},
  \bibinfo{pages}{A1133} (\bibinfo{year}{1965}).

\bibitem[{\citenamefont{Dobson et~al.}(1998)\citenamefont{Dobson, Vignale, and
  Das}}]{dftbook}
\bibinfo{author}{\bibfnamefont{J.~F.} \bibnamefont{Dobson}},
  \bibinfo{author}{\bibfnamefont{G.}~\bibnamefont{Vignale}}, \bibnamefont{and}
  \bibinfo{author}{\bibfnamefont{M.~P.} \bibnamefont{Das}},
  \emph{\bibinfo{title}{Electronic Density Functional Theory}}
  (\bibinfo{publisher}{Springer}, \bibinfo{year}{1998}).

\bibitem[{\citenamefont{Scuseria and Staroverov}(2005)}]{scus_rev}
\bibinfo{author}{\bibfnamefont{G.~E.} \bibnamefont{Scuseria}} \bibnamefont{and}
  \bibinfo{author}{\bibfnamefont{V.~N.} \bibnamefont{Staroverov}}, in
  \emph{\bibinfo{booktitle}{Theory and Applications of Computational Chemistry:
  The First 40 Years (A Volume of Technical and Historical Perspectives)}},
  edited by \bibinfo{editor}{\bibfnamefont{C.~E.} \bibnamefont{Dykstra}},
  \bibinfo{editor}{\bibfnamefont{G.}~\bibnamefont{Frenking}},
  \bibinfo{editor}{\bibfnamefont{K.~S.} \bibnamefont{Kim}}, \bibnamefont{and}
  \bibinfo{editor}{\bibfnamefont{G.~E.} \bibnamefont{Scuseria}}
  (\bibinfo{publisher}{Elsevier}, \bibinfo{address}{Amsterdam},
  \bibinfo{year}{2005}), pp. \bibinfo{pages}{669--724}.

\bibitem[{\citenamefont{Perdew and Schmidt}(2001)}]{perdewAIPCP2001}
\bibinfo{author}{\bibfnamefont{J.~P.} \bibnamefont{Perdew}} \bibnamefont{and}
  \bibinfo{author}{\bibfnamefont{K.}~\bibnamefont{Schmidt}},
  \bibinfo{journal}{AIP Conf. Proc.} \textbf{\bibinfo{volume}{577}},
  \bibinfo{pages}{1} (\bibinfo{year}{2001}).

\bibitem[{\citenamefont{Langreth and Mehl}(1983)}]{langrethPRB1983}
\bibinfo{author}{\bibfnamefont{D.~C.} \bibnamefont{Langreth}} \bibnamefont{and}
  \bibinfo{author}{\bibfnamefont{M.~J.} \bibnamefont{Mehl}},
  \bibinfo{journal}{Phys. Rev. B} \textbf{\bibinfo{volume}{28}},
  \bibinfo{pages}{1809} (\bibinfo{year}{1983}).

\bibitem[{\citenamefont{Perdew et~al.}(1996{\natexlab{a}})\citenamefont{Perdew,
  Ernzerhof, and Burke}}]{pbe0_2}
\bibinfo{author}{\bibfnamefont{J.~P.} \bibnamefont{Perdew}},
  \bibinfo{author}{\bibfnamefont{M.}~\bibnamefont{Ernzerhof}},
  \bibnamefont{and} \bibinfo{author}{\bibfnamefont{K.}~\bibnamefont{Burke}},
  \bibinfo{journal}{J. Chem. Phys.} \textbf{\bibinfo{volume}{105}},
  \bibinfo{pages}{9982} (\bibinfo{year}{1996}{\natexlab{a}}).

\bibitem[{\citenamefont{Arbuznikov}(2007)}]{arbuznikovJSC2007}
\bibinfo{author}{\bibfnamefont{A.~V.} \bibnamefont{Arbuznikov}},
  \bibinfo{journal}{J. Struct. Chem.} \textbf{\bibinfo{volume}{48}},
  \bibinfo{pages}{S1} (\bibinfo{year}{2007}).

\bibitem[{\citenamefont{Tao et~al.}(2003)\citenamefont{Tao, Perdew, Staroverov,
  and Scuseria}}]{tpss}
\bibinfo{author}{\bibfnamefont{J.}~\bibnamefont{Tao}},
  \bibinfo{author}{\bibfnamefont{J.~P.} \bibnamefont{Perdew}},
  \bibinfo{author}{\bibfnamefont{V.~N.} \bibnamefont{Staroverov}},
  \bibnamefont{and} \bibinfo{author}{\bibfnamefont{G.~E.}
  \bibnamefont{Scuseria}}, \bibinfo{journal}{Phys. Rev. Lett.}
  \textbf{\bibinfo{volume}{91}}, \bibinfo{pages}{146401}
  (\bibinfo{year}{2003}).

\bibitem[{\citenamefont{{Della Sala} et~al.}(2016)\citenamefont{{Della Sala},
  Fabiano, and Constantin}}]{kinrev}
\bibinfo{author}{\bibfnamefont{F.}~\bibnamefont{{Della Sala}}},
  \bibinfo{author}{\bibfnamefont{E.}~\bibnamefont{Fabiano}}, \bibnamefont{and}
  \bibinfo{author}{\bibfnamefont{L.~A.} \bibnamefont{Constantin}},
  \bibinfo{journal}{Int. J. Quantum Chem.} \textbf{\bibinfo{volume}{116}},
  \bibinfo{pages}{1641} (\bibinfo{year}{2016}).

\bibitem[{\citenamefont{Perdew et~al.}(2009)\citenamefont{Perdew, Ruzsinszky,
  Csonka, Constantin, and Sun}}]{revtpss}
\bibinfo{author}{\bibfnamefont{J.~P.} \bibnamefont{Perdew}},
  \bibinfo{author}{\bibfnamefont{A.}~\bibnamefont{Ruzsinszky}},
  \bibinfo{author}{\bibfnamefont{G.~I.} \bibnamefont{Csonka}},
  \bibinfo{author}{\bibfnamefont{L.~A.} \bibnamefont{Constantin}},
  \bibnamefont{and} \bibinfo{author}{\bibfnamefont{J.}~\bibnamefont{Sun}},
  \bibinfo{journal}{Phys. Rev. Lett.} \textbf{\bibinfo{volume}{103}},
  \bibinfo{pages}{026403} (\bibinfo{year}{2009}).

\bibitem[{\citenamefont{Constantin
  et~al.}(2011{\natexlab{a}})\citenamefont{Constantin, Chiodo, Fabiano,
  Bodrenko, and {Della Sala}}}]{js}
\bibinfo{author}{\bibfnamefont{L.~A.} \bibnamefont{Constantin}},
  \bibinfo{author}{\bibfnamefont{L.}~\bibnamefont{Chiodo}},
  \bibinfo{author}{\bibfnamefont{E.}~\bibnamefont{Fabiano}},
  \bibinfo{author}{\bibfnamefont{I.}~\bibnamefont{Bodrenko}}, \bibnamefont{and}
  \bibinfo{author}{\bibfnamefont{F.}~\bibnamefont{{Della Sala}}},
  \bibinfo{journal}{Phys. Rev. B} \textbf{\bibinfo{volume}{84}},
  \bibinfo{pages}{045126} (\bibinfo{year}{2011}{\natexlab{a}}).

\bibitem[{\citenamefont{Constantin et~al.}(2012)\citenamefont{Constantin,
  Fabiano, and {Della Sala}}}]{tpssloc}
\bibinfo{author}{\bibfnamefont{L.~A.} \bibnamefont{Constantin}},
  \bibinfo{author}{\bibfnamefont{E.}~\bibnamefont{Fabiano}}, \bibnamefont{and}
  \bibinfo{author}{\bibfnamefont{F.}~\bibnamefont{{Della Sala}}},
  \bibinfo{journal}{Phys. Rev. B} \textbf{\bibinfo{volume}{86}},
  \bibinfo{pages}{035130} (\bibinfo{year}{2012}).

\bibitem[{\citenamefont{Constantin
  et~al.}(2013{\natexlab{a}})\citenamefont{Constantin, Fabiano, and {Della
  Sala}}}]{bloc_hole}
\bibinfo{author}{\bibfnamefont{L.~A.} \bibnamefont{Constantin}},
  \bibinfo{author}{\bibfnamefont{E.}~\bibnamefont{Fabiano}}, \bibnamefont{and}
  \bibinfo{author}{\bibfnamefont{F.}~\bibnamefont{{Della Sala}}},
  \bibinfo{journal}{Phys. Rev. B} \textbf{\bibinfo{volume}{88}},
  \bibinfo{pages}{125112} (\bibinfo{year}{2013}{\natexlab{a}}).

\bibitem[{\citenamefont{{Della Sala} et~al.}(2015)\citenamefont{{Della Sala},
  Fabiano, and Constantin}}]{alpha}
\bibinfo{author}{\bibfnamefont{F.}~\bibnamefont{{Della Sala}}},
  \bibinfo{author}{\bibfnamefont{E.}~\bibnamefont{Fabiano}}, \bibnamefont{and}
  \bibinfo{author}{\bibfnamefont{L.~A.} \bibnamefont{Constantin}},
  \bibinfo{journal}{Phys. Rev. B} \textbf{\bibinfo{volume}{91}},
  \bibinfo{pages}{035126} (\bibinfo{year}{2015}).

\bibitem[{\citenamefont{Van~Voorhis and Scuseria}(1998)}]{vsxc}
\bibinfo{author}{\bibfnamefont{T.}~\bibnamefont{Van~Voorhis}} \bibnamefont{and}
  \bibinfo{author}{\bibfnamefont{G.~E.} \bibnamefont{Scuseria}},
  \bibinfo{journal}{J. Chem. Phys.} \textbf{\bibinfo{volume}{109}},
  \bibinfo{pages}{400} (\bibinfo{year}{1998}).

\bibitem[{\citenamefont{Schmider and Becke}(1998)}]{schmider98}
\bibinfo{author}{\bibfnamefont{H.~L.} \bibnamefont{Schmider}} \bibnamefont{and}
  \bibinfo{author}{\bibfnamefont{A.~D.} \bibnamefont{Becke}},
  \bibinfo{journal}{J. Chem. Phys.} \textbf{\bibinfo{volume}{109}},
  \bibinfo{pages}{8188} (\bibinfo{year}{1998}).

\bibitem[{\citenamefont{Zhao and Truhlar}(2006)}]{m06l}
\bibinfo{author}{\bibfnamefont{Y.}~\bibnamefont{Zhao}} \bibnamefont{and}
  \bibinfo{author}{\bibfnamefont{D.~G.} \bibnamefont{Truhlar}},
  \bibinfo{journal}{J. Chem. Phys.} \textbf{\bibinfo{volume}{125}},
  \bibinfo{eid}{194101} (\bibinfo{year}{2006}).

\bibitem[{\citenamefont{Ruzsinszky et~al.}(2012)\citenamefont{Ruzsinszky, Sun,
  Xiao, and Csonka}}]{regtpss}
\bibinfo{author}{\bibfnamefont{A.}~\bibnamefont{Ruzsinszky}},
  \bibinfo{author}{\bibfnamefont{J.}~\bibnamefont{Sun}},
  \bibinfo{author}{\bibfnamefont{B.}~\bibnamefont{Xiao}}, \bibnamefont{and}
  \bibinfo{author}{\bibfnamefont{G.~I.} \bibnamefont{Csonka}},
  \bibinfo{journal}{J. Chem. Theory. Comput.} \textbf{\bibinfo{volume}{8}},
  \bibinfo{pages}{2078} (\bibinfo{year}{2012}).

\bibitem[{\citenamefont{Sun et~al.}(2012)\citenamefont{Sun, Xiao, and
  Ruzsinszky}}]{mggms_1}
\bibinfo{author}{\bibfnamefont{J.}~\bibnamefont{Sun}},
  \bibinfo{author}{\bibfnamefont{B.}~\bibnamefont{Xiao}}, \bibnamefont{and}
  \bibinfo{author}{\bibfnamefont{A.}~\bibnamefont{Ruzsinszky}},
  \bibinfo{journal}{J. Chem. Phys.} \textbf{\bibinfo{volume}{137}},
  \bibinfo{eid}{051101} (\bibinfo{year}{2012}).

\bibitem[{\citenamefont{Sun et~al.}(2013{\natexlab{a}})\citenamefont{Sun,
  Haunschild, Xiao, Bulik, Scuseria, and Perdew}}]{mgga_ms2}
\bibinfo{author}{\bibfnamefont{J.}~\bibnamefont{Sun}},
  \bibinfo{author}{\bibfnamefont{R.}~\bibnamefont{Haunschild}},
  \bibinfo{author}{\bibfnamefont{B.}~\bibnamefont{Xiao}},
  \bibinfo{author}{\bibfnamefont{I.~W.} \bibnamefont{Bulik}},
  \bibinfo{author}{\bibfnamefont{G.~E.} \bibnamefont{Scuseria}},
  \bibnamefont{and} \bibinfo{author}{\bibfnamefont{J.~P.}
  \bibnamefont{Perdew}}, \bibinfo{journal}{J. Chem. Phys.}
  \textbf{\bibinfo{volume}{138}}, \bibinfo{eid}{044113}
  (\bibinfo{year}{2013}{\natexlab{a}}).

\bibitem[{\citenamefont{Peverati and Truhlar}(2012)}]{m11l}
\bibinfo{author}{\bibfnamefont{R.}~\bibnamefont{Peverati}} \bibnamefont{and}
  \bibinfo{author}{\bibfnamefont{D.~G.} \bibnamefont{Truhlar}},
  \bibinfo{journal}{J. Phys. Chem. Letters} \textbf{\bibinfo{volume}{3}},
  \bibinfo{pages}{117} (\bibinfo{year}{2012}).

\bibitem[{\citenamefont{Xiao et~al.}(2013)\citenamefont{Xiao, Sun, Ruzsinszky,
  Feng, Haunschild, Scuseria, and Perdew}}]{xiaosolid}
\bibinfo{author}{\bibfnamefont{B.}~\bibnamefont{Xiao}},
  \bibinfo{author}{\bibfnamefont{J.}~\bibnamefont{Sun}},
  \bibinfo{author}{\bibfnamefont{A.}~\bibnamefont{Ruzsinszky}},
  \bibinfo{author}{\bibfnamefont{J.}~\bibnamefont{Feng}},
  \bibinfo{author}{\bibfnamefont{R.}~\bibnamefont{Haunschild}},
  \bibinfo{author}{\bibfnamefont{G.~E.} \bibnamefont{Scuseria}},
  \bibnamefont{and} \bibinfo{author}{\bibfnamefont{J.~P.}
  \bibnamefont{Perdew}}, \bibinfo{journal}{Phys. Rev. B}
  \textbf{\bibinfo{volume}{88}}, \bibinfo{pages}{184103}
  (\bibinfo{year}{2013}).

\bibitem[{\citenamefont{Sun et~al.}(2013{\natexlab{b}})\citenamefont{Sun, Xiao,
  Fang, Haunschild, Hao, Ruzsinszky, Csonka, Scuseria, and Perdew}}]{sun13}
\bibinfo{author}{\bibfnamefont{J.}~\bibnamefont{Sun}},
  \bibinfo{author}{\bibfnamefont{B.}~\bibnamefont{Xiao}},
  \bibinfo{author}{\bibfnamefont{Y.}~\bibnamefont{Fang}},
  \bibinfo{author}{\bibfnamefont{R.}~\bibnamefont{Haunschild}},
  \bibinfo{author}{\bibfnamefont{P.}~\bibnamefont{Hao}},
  \bibinfo{author}{\bibfnamefont{A.}~\bibnamefont{Ruzsinszky}},
  \bibinfo{author}{\bibfnamefont{G.~I.} \bibnamefont{Csonka}},
  \bibinfo{author}{\bibfnamefont{G.~E.} \bibnamefont{Scuseria}},
  \bibnamefont{and} \bibinfo{author}{\bibfnamefont{J.~P.}
  \bibnamefont{Perdew}}, \bibinfo{journal}{Phys. Rev. Lett.}
  \textbf{\bibinfo{volume}{111}}, \bibinfo{pages}{106401}
  (\bibinfo{year}{2013}{\natexlab{b}}).

\bibitem[{\citenamefont{Staroverov et~al.}(2004)\citenamefont{Staroverov,
  Scuseria, Tao, and Perdew}}]{stare04}
\bibinfo{author}{\bibfnamefont{V.~N.} \bibnamefont{Staroverov}},
  \bibinfo{author}{\bibfnamefont{G.~E.} \bibnamefont{Scuseria}},
  \bibinfo{author}{\bibfnamefont{J.}~\bibnamefont{Tao}}, \bibnamefont{and}
  \bibinfo{author}{\bibfnamefont{J.~P.} \bibnamefont{Perdew}},
  \bibinfo{journal}{Phys. Rev. B} \textbf{\bibinfo{volume}{69}},
  \bibinfo{pages}{075102} (\bibinfo{year}{2004}).

\bibitem[{\citenamefont{Adamo et~al.}(2000)\citenamefont{Adamo, Ernzerhof, and
  Scuseria}}]{adamo00}
\bibinfo{author}{\bibfnamefont{C.}~\bibnamefont{Adamo}},
  \bibinfo{author}{\bibfnamefont{M.}~\bibnamefont{Ernzerhof}},
  \bibnamefont{and} \bibinfo{author}{\bibfnamefont{G.~E.}
  \bibnamefont{Scuseria}}, \bibinfo{journal}{J. Chem. Phys.}
  \textbf{\bibinfo{volume}{112}}, \bibinfo{pages}{2643} (\bibinfo{year}{2000}).

\bibitem[{\citenamefont{Riley et~al.}(2007)\citenamefont{Riley, Op't~Holt, and
  Merz}}]{riley07}
\bibinfo{author}{\bibfnamefont{K.~E.} \bibnamefont{Riley}},
  \bibinfo{author}{\bibfnamefont{B.~T.} \bibnamefont{Op't~Holt}},
  \bibnamefont{and} \bibinfo{author}{\bibfnamefont{K.~M.} \bibnamefont{Merz}},
  \bibinfo{journal}{J. Chem. Theory. Comput.} \textbf{\bibinfo{volume}{3}},
  \bibinfo{pages}{407} (\bibinfo{year}{2007}).

\bibitem[{\citenamefont{Sun et~al.}(2015)\citenamefont{Sun, Ruzsinszky, and
  Perdew}}]{scan}
\bibinfo{author}{\bibfnamefont{J.}~\bibnamefont{Sun}},
  \bibinfo{author}{\bibfnamefont{A.}~\bibnamefont{Ruzsinszky}},
  \bibnamefont{and} \bibinfo{author}{\bibfnamefont{J.~P.}
  \bibnamefont{Perdew}}, \bibinfo{journal}{Phys. Rev. Lett.}
  \textbf{\bibinfo{volume}{115}}, \bibinfo{pages}{036402}
  (\bibinfo{year}{2015}).

\bibitem[{\citenamefont{Sun et~al.}(2016)\citenamefont{Sun, Remsing, Zhang,
  Sun, Ruzsinszky, Peng, Yang, Paul, Waghmare, Wu et~al.}}]{sunNC2016}
\bibinfo{author}{\bibfnamefont{J.}~\bibnamefont{Sun}},
  \bibinfo{author}{\bibfnamefont{R.~C.} \bibnamefont{Remsing}},
  \bibinfo{author}{\bibfnamefont{Y.}~\bibnamefont{Zhang}},
  \bibinfo{author}{\bibfnamefont{Z.}~\bibnamefont{Sun}},
  \bibinfo{author}{\bibfnamefont{A.}~\bibnamefont{Ruzsinszky}},
  \bibinfo{author}{\bibfnamefont{H.}~\bibnamefont{Peng}},
  \bibinfo{author}{\bibfnamefont{Z.}~\bibnamefont{Yang}},
  \bibinfo{author}{\bibfnamefont{A.}~\bibnamefont{Paul}},
  \bibinfo{author}{\bibfnamefont{U.}~\bibnamefont{Waghmare}},
  \bibinfo{author}{\bibfnamefont{X.}~\bibnamefont{Wu}}, \bibnamefont{et~al.},
  \bibinfo{journal}{Nat. Chem.} \textbf{\bibinfo{volume}{8}},
  \bibinfo{pages}{831} (\bibinfo{year}{2016}), \bibinfo{note}{article}.

\bibitem[{\citenamefont{Yu et~al.}(2016)\citenamefont{Yu, He, and
  Truhlar}}]{m15l}
\bibinfo{author}{\bibfnamefont{H.~S.} \bibnamefont{Yu}},
  \bibinfo{author}{\bibfnamefont{X.}~\bibnamefont{He}}, \bibnamefont{and}
  \bibinfo{author}{\bibfnamefont{D.~G.} \bibnamefont{Truhlar}},
  \bibinfo{journal}{J. Chem. Theory. Comput.} \textbf{\bibinfo{volume}{12}},
  \bibinfo{pages}{1280} (\bibinfo{year}{2016}).

\bibitem[{\citenamefont{Tao and Mo}(2016)}]{tao16}
\bibinfo{author}{\bibfnamefont{J.}~\bibnamefont{Tao}} \bibnamefont{and}
  \bibinfo{author}{\bibfnamefont{Y.}~\bibnamefont{Mo}}, \bibinfo{journal}{Phys.
  Rev. Lett.} \textbf{\bibinfo{volume}{117}}, \bibinfo{pages}{073001}
  (\bibinfo{year}{2016}).

\bibitem[{\citenamefont{Wellendorff et~al.}(2014)\citenamefont{Wellendorff,
  Lundgaard, Jacobsen, and Bligaard}}]{mbeef_2014}
\bibinfo{author}{\bibfnamefont{J.}~\bibnamefont{Wellendorff}},
  \bibinfo{author}{\bibfnamefont{K.~T.} \bibnamefont{Lundgaard}},
  \bibinfo{author}{\bibfnamefont{K.~W.} \bibnamefont{Jacobsen}},
  \bibnamefont{and} \bibinfo{author}{\bibfnamefont{T.}~\bibnamefont{Bligaard}},
  \bibinfo{journal}{J. Chem. Phys.} \textbf{\bibinfo{volume}{140}},
  \bibinfo{eid}{144107} (\bibinfo{year}{2014}).

\bibitem[{\citenamefont{Mardirossian and Head-Gordon}(2015)}]{b97mv_2015}
\bibinfo{author}{\bibfnamefont{N.}~\bibnamefont{Mardirossian}}
  \bibnamefont{and}
  \bibinfo{author}{\bibfnamefont{M.}~\bibnamefont{Head-Gordon}},
  \bibinfo{journal}{J. Chem. Phys.} \textbf{\bibinfo{volume}{142}},
  \bibinfo{eid}{074111} (pages~\bibinfo{numpages}{1}) (\bibinfo{year}{2015}).

\bibitem[{\citenamefont{Constantin et~al.}(2016)\citenamefont{Constantin,
  Fabiano, Pitarke, and Della~Sala}}]{satpss_2016}
\bibinfo{author}{\bibfnamefont{L.~A.} \bibnamefont{Constantin}},
  \bibinfo{author}{\bibfnamefont{E.}~\bibnamefont{Fabiano}},
  \bibinfo{author}{\bibfnamefont{J.}~\bibnamefont{Pitarke}}, \bibnamefont{and}
  \bibinfo{author}{\bibfnamefont{F.}~\bibnamefont{Della~Sala}},
  \bibinfo{journal}{Phys. Rev. B} \textbf{\bibinfo{volume}{93}},
  \bibinfo{pages}{115127} (\bibinfo{year}{2016}).

\bibitem[{\citenamefont{K\"ummel and Kronik}(2008)}]{kummel2008}
\bibinfo{author}{\bibfnamefont{S.}~\bibnamefont{K\"ummel}} \bibnamefont{and}
  \bibinfo{author}{\bibfnamefont{L.}~\bibnamefont{Kronik}},
  \bibinfo{journal}{Rev. Mod. Phys.} \textbf{\bibinfo{volume}{80}},
  \bibinfo{pages}{3} (\bibinfo{year}{2008}).

\bibitem[{\citenamefont{Arbuznikov and Kaupp}(2003)}]{arbuznikov03}
\bibinfo{author}{\bibfnamefont{A.~V.} \bibnamefont{Arbuznikov}}
  \bibnamefont{and} \bibinfo{author}{\bibfnamefont{M.}~\bibnamefont{Kaupp}},
  \bibinfo{journal}{Chem. Phys. Lett.} \textbf{\bibinfo{volume}{381}},
  \bibinfo{pages}{495 } (\bibinfo{year}{2003}).

\bibitem[{\citenamefont{Zahariev et~al.}(2013)\citenamefont{Zahariev, Leang,
  and Gordon}}]{zaharievJCP2013}
\bibinfo{author}{\bibfnamefont{F.}~\bibnamefont{Zahariev}},
  \bibinfo{author}{\bibfnamefont{S.~S.} \bibnamefont{Leang}}, \bibnamefont{and}
  \bibinfo{author}{\bibfnamefont{M.~S.} \bibnamefont{Gordon}},
  \bibinfo{journal}{J. Chem. Phys.} \textbf{\bibinfo{volume}{138}},
  \bibinfo{eid}{244108} (\bibinfo{year}{2013}).

\bibitem[{\citenamefont{Seidl et~al.}(1996)\citenamefont{Seidl, G\"orling,
  Vogl, Majewski, and Levy}}]{seidl96}
\bibinfo{author}{\bibfnamefont{A.}~\bibnamefont{Seidl}},
  \bibinfo{author}{\bibfnamefont{A.}~\bibnamefont{G\"orling}},
  \bibinfo{author}{\bibfnamefont{P.}~\bibnamefont{Vogl}},
  \bibinfo{author}{\bibfnamefont{J.~A.} \bibnamefont{Majewski}},
  \bibnamefont{and} \bibinfo{author}{\bibfnamefont{M.}~\bibnamefont{Levy}},
  \bibinfo{journal}{Phys. Rev. B} \textbf{\bibinfo{volume}{53}},
  \bibinfo{pages}{3764} (\bibinfo{year}{1996}).

\bibitem[{\citenamefont{Wesolowski}(2006)}]{wesorev}
\bibinfo{author}{\bibfnamefont{T.~A.} \bibnamefont{Wesolowski}}, in
  \emph{\bibinfo{booktitle}{Chemistry: Reviews of Current Trends}}, edited by
  \bibinfo{editor}{\bibfnamefont{J.}~\bibnamefont{Leszczynski}}
  (\bibinfo{publisher}{World Scientific: Singapore, 2006},
  \bibinfo{address}{Singapore}, \bibinfo{year}{2006}),
  vol.~\bibinfo{volume}{10}, pp. \bibinfo{pages}{1--82}.

\bibitem[{\citenamefont{Jacob and Neugebauer}(2014)}]{subdft_rev}
\bibinfo{author}{\bibfnamefont{C.~R.} \bibnamefont{Jacob}} \bibnamefont{and}
  \bibinfo{author}{\bibfnamefont{J.}~\bibnamefont{Neugebauer}},
  \bibinfo{journal}{Wiley Interdisciplinary Reviews: Computational Molecular
  Science} \textbf{\bibinfo{volume}{4}}, \bibinfo{pages}{325}
  (\bibinfo{year}{2014}).

\bibitem[{\citenamefont{Krishtal et~al.}(2015)\citenamefont{Krishtal, Sinha,
  Genova, and Pavanello}}]{krishtal15}
\bibinfo{author}{\bibfnamefont{A.}~\bibnamefont{Krishtal}},
  \bibinfo{author}{\bibfnamefont{D.}~\bibnamefont{Sinha}},
  \bibinfo{author}{\bibfnamefont{A.}~\bibnamefont{Genova}}, \bibnamefont{and}
  \bibinfo{author}{\bibfnamefont{M.}~\bibnamefont{Pavanello}},
  \bibinfo{journal}{J. Phys.-Condens. Mat.} \textbf{\bibinfo{volume}{27}},
  \bibinfo{pages}{183202} (\bibinfo{year}{2015}).

\bibitem[{\citenamefont{Prodan and Kohn}(2005)}]{prodan}
\bibinfo{author}{\bibfnamefont{E.}~\bibnamefont{Prodan}} \bibnamefont{and}
  \bibinfo{author}{\bibfnamefont{W.}~\bibnamefont{Kohn}},
  \bibinfo{journal}{PNAS} \textbf{\bibinfo{volume}{102}},
  \bibinfo{pages}{11635} (\bibinfo{year}{2005}).

\bibitem[{\citenamefont{Wesolowski and Warshel}(1993)}]{wesowarh93}
\bibinfo{author}{\bibfnamefont{T.~A.} \bibnamefont{Wesolowski}}
  \bibnamefont{and} \bibinfo{author}{\bibfnamefont{A.}~\bibnamefont{Warshel}},
  \bibinfo{journal}{J. Phys. Chem.} \textbf{\bibinfo{volume}{97}},
  \bibinfo{pages}{8050} (\bibinfo{year}{1993}).

\bibitem[{\citenamefont{Wesolowski and Weber}(1996)}]{Wesolowski199671}
\bibinfo{author}{\bibfnamefont{T.~A.} \bibnamefont{Wesolowski}}
  \bibnamefont{and} \bibinfo{author}{\bibfnamefont{J.}~\bibnamefont{Weber}},
  \bibinfo{journal}{Chem. Phys. Lett.} \textbf{\bibinfo{volume}{248}},
  \bibinfo{pages}{71 } (\bibinfo{year}{1996}).

\bibitem[{\citenamefont{G\"{o}tz et~al.}(2009)\citenamefont{G\"{o}tz, Beyhan,
  and Visscher}}]{gotz09}
\bibinfo{author}{\bibfnamefont{A.~W.} \bibnamefont{G\"{o}tz}},
  \bibinfo{author}{\bibfnamefont{S.~M.} \bibnamefont{Beyhan}},
  \bibnamefont{and} \bibinfo{author}{\bibfnamefont{L.}~\bibnamefont{Visscher}},
  \bibinfo{journal}{J. Chem. Theory Comput.} \textbf{\bibinfo{volume}{5}},
  \bibinfo{pages}{3161} (\bibinfo{year}{2009}).

\bibitem[{\citenamefont{Wesolowski}(1997)}]{wesolowski97hyd}
\bibinfo{author}{\bibfnamefont{T.~A.} \bibnamefont{Wesolowski}},
  \bibinfo{journal}{J. Chem. Phys.} \textbf{\bibinfo{volume}{106}},
  \bibinfo{pages}{8516} (\bibinfo{year}{1997}).

\bibitem[{\citenamefont{Wesolowski et~al.}(1996)\citenamefont{Wesolowski,
  Chermette, and Weber}}]{wesolowski96fhnch}
\bibinfo{author}{\bibfnamefont{T.~A.} \bibnamefont{Wesolowski}},
  \bibinfo{author}{\bibfnamefont{H.}~\bibnamefont{Chermette}},
  \bibnamefont{and} \bibinfo{author}{\bibfnamefont{J.}~\bibnamefont{Weber}},
  \bibinfo{journal}{J. Chem. Phys.} \textbf{\bibinfo{volume}{105}},
  \bibinfo{pages}{9182} (\bibinfo{year}{1996}).

\bibitem[{\citenamefont{Constantin
  et~al.}(2011{\natexlab{b}})\citenamefont{Constantin, Fabiano, Laricchia, and
  {Della Sala}}}]{apbe}
\bibinfo{author}{\bibfnamefont{L.~A.} \bibnamefont{Constantin}},
  \bibinfo{author}{\bibfnamefont{E.}~\bibnamefont{Fabiano}},
  \bibinfo{author}{\bibfnamefont{S.}~\bibnamefont{Laricchia}},
  \bibnamefont{and} \bibinfo{author}{\bibfnamefont{F.}~\bibnamefont{{Della
  Sala}}}, \bibinfo{journal}{Phys. Rev. Lett.} \textbf{\bibinfo{volume}{106}},
  \bibinfo{pages}{186406} (\bibinfo{year}{2011}{\natexlab{b}}).

\bibitem[{\citenamefont{Laricchia
  et~al.}(2011{\natexlab{a}})\citenamefont{Laricchia, Fabiano, Constantin, and
  {Della Sala}}}]{apbek}
\bibinfo{author}{\bibfnamefont{S.}~\bibnamefont{Laricchia}},
  \bibinfo{author}{\bibfnamefont{E.}~\bibnamefont{Fabiano}},
  \bibinfo{author}{\bibfnamefont{L.~A.} \bibnamefont{Constantin}},
  \bibnamefont{and} \bibinfo{author}{\bibfnamefont{F.}~\bibnamefont{{Della
  Sala}}}, \bibinfo{journal}{J. Chem. Theory. Comput.}
  \textbf{\bibinfo{volume}{7}}, \bibinfo{pages}{2439}
  (\bibinfo{year}{2011}{\natexlab{a}}).

\bibitem[{\citenamefont{Laricchia et~al.}(2014)\citenamefont{Laricchia,
  Constantin, Fabiano, and {Della Sala}}}]{fde_lap}
\bibinfo{author}{\bibfnamefont{S.}~\bibnamefont{Laricchia}},
  \bibinfo{author}{\bibfnamefont{L.~A.} \bibnamefont{Constantin}},
  \bibinfo{author}{\bibfnamefont{E.}~\bibnamefont{Fabiano}}, \bibnamefont{and}
  \bibinfo{author}{\bibfnamefont{F.}~\bibnamefont{{Della Sala}}},
  \bibinfo{journal}{J. Chem. Theory. Comput.} \textbf{\bibinfo{volume}{10}},
  \bibinfo{pages}{164} (\bibinfo{year}{2014}).

\bibitem[{\citenamefont{Tran and Wesołowski}(2002)}]{tran02link}
\bibinfo{author}{\bibfnamefont{F.}~\bibnamefont{Tran}} \bibnamefont{and}
  \bibinfo{author}{\bibfnamefont{T.~A.} \bibnamefont{Wesołowski}},
  \bibinfo{journal}{Int. J. Quantum Chem.} \textbf{\bibinfo{volume}{89}},
  \bibinfo{pages}{441} (\bibinfo{year}{2002}).

\bibitem[{\citenamefont{Tran and Wesolowski}(2013)}]{weso_chap}
\bibinfo{author}{\bibfnamefont{F.}~\bibnamefont{Tran}} \bibnamefont{and}
  \bibinfo{author}{\bibfnamefont{T.~A.} \bibnamefont{Wesolowski}}, in
  \emph{\bibinfo{booktitle}{Recent Progress in Orbital-free Density Functional
  Theory}}, edited by \bibinfo{editor}{\bibfnamefont{T.~A.}
  \bibnamefont{Wesolowsky}} \bibnamefont{and}
  \bibinfo{editor}{\bibfnamefont{Y.~A.} \bibnamefont{Wang}}
  (\bibinfo{publisher}{World Scientific}, \bibinfo{address}{Singapore},
  \bibinfo{year}{2013}), pp. \bibinfo{pages}{429--442}.

\bibitem[{\citenamefont{Fabiano
  et~al.}(2014{\natexlab{a}})\citenamefont{Fabiano, Laricchia, and {Della
  Sala}}}]{fde_fractional}
\bibinfo{author}{\bibfnamefont{E.}~\bibnamefont{Fabiano}},
  \bibinfo{author}{\bibfnamefont{S.}~\bibnamefont{Laricchia}},
  \bibnamefont{and} \bibinfo{author}{\bibfnamefont{F.}~\bibnamefont{{Della
  Sala}}}, \bibinfo{journal}{J. Chem. Phys.} \textbf{\bibinfo{volume}{140}},
  \bibinfo{eid}{114101} (\bibinfo{year}{2014}{\natexlab{a}}).

\bibitem[{\citenamefont{Laricchia
  et~al.}(2011{\natexlab{b}})\citenamefont{Laricchia, Fabiano, and {Della
  Sala}}}]{fde_lhf}
\bibinfo{author}{\bibfnamefont{S.}~\bibnamefont{Laricchia}},
  \bibinfo{author}{\bibfnamefont{E.}~\bibnamefont{Fabiano}}, \bibnamefont{and}
  \bibinfo{author}{\bibfnamefont{F.}~\bibnamefont{{Della Sala}}},
  \bibinfo{journal}{Chem. Phys. Lett.} \textbf{\bibinfo{volume}{518}},
  \bibinfo{pages}{114 } (\bibinfo{year}{2011}{\natexlab{b}}).

\bibitem[{\citenamefont{Weso\l{}owski}(2008)}]{wesolowski08}
\bibinfo{author}{\bibfnamefont{T.~A.} \bibnamefont{Weso\l{}owski}},
  \bibinfo{journal}{Phys. Rev. A} \textbf{\bibinfo{volume}{77}},
  \bibinfo{pages}{012504} (\bibinfo{year}{2008}).

\bibitem[{\citenamefont{Dresselhaus and Neugebauer}(2015)}]{Dresselhaus2015}
\bibinfo{author}{\bibfnamefont{T.}~\bibnamefont{Dresselhaus}} \bibnamefont{and}
  \bibinfo{author}{\bibfnamefont{J.}~\bibnamefont{Neugebauer}},
  \bibinfo{journal}{Theor. Chem. Acc.} \textbf{\bibinfo{volume}{134}},
  \bibinfo{pages}{1} (\bibinfo{year}{2015}).

\bibitem[{\citenamefont{Laricchia et~al.}(2010)\citenamefont{Laricchia,
  Fabiano, and {Della Sala}}}]{fde_hybrid}
\bibinfo{author}{\bibfnamefont{S.}~\bibnamefont{Laricchia}},
  \bibinfo{author}{\bibfnamefont{E.}~\bibnamefont{Fabiano}}, \bibnamefont{and}
  \bibinfo{author}{\bibfnamefont{F.}~\bibnamefont{{Della Sala}}},
  \bibinfo{journal}{J. Chem. Phys.} \textbf{\bibinfo{volume}{133}},
  \bibinfo{eid}{164111} (\bibinfo{year}{2010}).

\bibitem[{\citenamefont{\'Smiga et~al.}(2015)\citenamefont{\'Smiga, Fabiano,
  Laricchia, Constantin, and {Della Sala}}}]{fdemeta}
\bibinfo{author}{\bibfnamefont{S.}~\bibnamefont{\'Smiga}},
  \bibinfo{author}{\bibfnamefont{E.}~\bibnamefont{Fabiano}},
  \bibinfo{author}{\bibfnamefont{S.}~\bibnamefont{Laricchia}},
  \bibinfo{author}{\bibfnamefont{L.~A.} \bibnamefont{Constantin}},
  \bibnamefont{and} \bibinfo{author}{\bibfnamefont{F.}~\bibnamefont{{Della
  Sala}}}, \bibinfo{journal}{J. Chem. Phys.} \textbf{\bibinfo{volume}{142}},
  \bibinfo{eid}{154121} (\bibinfo{year}{2015}).

\bibitem[{\citenamefont{Ramos et~al.}(2015)\citenamefont{Ramos, Papadakis, and
  Pavanello}}]{ramos15}
\bibinfo{author}{\bibfnamefont{P.}~\bibnamefont{Ramos}},
  \bibinfo{author}{\bibfnamefont{M.}~\bibnamefont{Papadakis}},
  \bibnamefont{and}
  \bibinfo{author}{\bibfnamefont{M.}~\bibnamefont{Pavanello}},
  \bibinfo{journal}{J. Phys. Chem. B} \textbf{\bibinfo{volume}{119}},
  \bibinfo{pages}{7541} (\bibinfo{year}{2015}).

\bibitem[{\citenamefont{Thomas}(1926)}]{thomas26}
\bibinfo{author}{\bibfnamefont{L.~H.} \bibnamefont{Thomas}},
  \bibinfo{journal}{Proc. Cambridge Phil. Soc.} \textbf{\bibinfo{volume}{23}},
  \bibinfo{pages}{542} (\bibinfo{year}{1926}).

\bibitem[{\citenamefont{Fermi}(1928)}]{fermi28}
\bibinfo{author}{\bibfnamefont{E.}~\bibnamefont{Fermi}},
  \bibinfo{journal}{Rend. Accad. Naz. Lincei} \textbf{\bibinfo{volume}{48}},
  \bibinfo{pages}{73} (\bibinfo{year}{1928}).

\bibitem[{\citenamefont{Fermi}(1927)}]{fermi27}
\bibinfo{author}{\bibfnamefont{E.}~\bibnamefont{Fermi}}, \bibinfo{journal}{Z.
  Phys.} \textbf{\bibinfo{volume}{6}}, \bibinfo{pages}{602}
  (\bibinfo{year}{1927}).

\bibitem[{\citenamefont{{von Weizs\"{a}cker}}(1935)}]{vw}
\bibinfo{author}{\bibfnamefont{C.~F.} \bibnamefont{{von Weizs\"{a}cker}}},
  \bibinfo{journal}{Z. Phys. A} \textbf{\bibinfo{volume}{96}},
  \bibinfo{pages}{431} (\bibinfo{year}{1935}).

\bibitem[{\citenamefont{Cirac\`{\i} and {Della Sala}}(2016)}]{QHD16}
\bibinfo{author}{\bibfnamefont{C.}~\bibnamefont{Cirac\`{\i}}} \bibnamefont{and}
  \bibinfo{author}{\bibfnamefont{F.}~\bibnamefont{{Della Sala}}},
  \bibinfo{journal}{Phys. Rev. B} \textbf{\bibinfo{volume}{93}},
  \bibinfo{pages}{205405} (\bibinfo{year}{2016}).

\bibitem[{\citenamefont{Kirzhnitz}(1967)}]{kirzhnitz1967field}
\bibinfo{author}{\bibfnamefont{D.~A.} \bibnamefont{Kirzhnitz}},
  \emph{\bibinfo{title}{Field theoretical methods in many-body systems}}
  (\bibinfo{publisher}{Pergamon Press}, \bibinfo{year}{1967}).

\bibitem[{\citenamefont{Brack et~al.}(1976)\citenamefont{Brack, Jennings, and
  Chu}}]{gea2}
\bibinfo{author}{\bibfnamefont{M.}~\bibnamefont{Brack}},
  \bibinfo{author}{\bibfnamefont{B.}~\bibnamefont{Jennings}}, \bibnamefont{and}
  \bibinfo{author}{\bibfnamefont{Y.}~\bibnamefont{Chu}},
  \bibinfo{journal}{Phys. Lett. B} \textbf{\bibinfo{volume}{65}},
  \bibinfo{pages}{1 } (\bibinfo{year}{1976}).

\bibitem[{\citenamefont{Yang}(1986)}]{yang1986gradient}
\bibinfo{author}{\bibfnamefont{W.}~\bibnamefont{Yang}}, \bibinfo{journal}{Phys.
  Rev. A} \textbf{\bibinfo{volume}{34}}, \bibinfo{pages}{4575}
  (\bibinfo{year}{1986}).

\bibitem[{\citenamefont{Yang et~al.}(1986)\citenamefont{Yang, Parr, and
  Lee}}]{yang86}
\bibinfo{author}{\bibfnamefont{W.}~\bibnamefont{Yang}},
  \bibinfo{author}{\bibfnamefont{R.~G.} \bibnamefont{Parr}}, \bibnamefont{and}
  \bibinfo{author}{\bibfnamefont{C.}~\bibnamefont{Lee}},
  \bibinfo{journal}{Phys. Rev. A} \textbf{\bibinfo{volume}{34}},
  \bibinfo{pages}{4586} (\bibinfo{year}{1986}).

\bibitem[{\citenamefont{Lee et~al.}(2009)\citenamefont{Lee, Constantin, Perdew,
  and Burke}}]{mge2}
\bibinfo{author}{\bibfnamefont{D.}~\bibnamefont{Lee}},
  \bibinfo{author}{\bibfnamefont{L.~A.} \bibnamefont{Constantin}},
  \bibinfo{author}{\bibfnamefont{J.~P.} \bibnamefont{Perdew}},
  \bibnamefont{and} \bibinfo{author}{\bibfnamefont{K.}~\bibnamefont{Burke}},
  \bibinfo{journal}{J. Chem. Phys.} \textbf{\bibinfo{volume}{130}},
  \bibinfo{eid}{034107} (\bibinfo{year}{2009}).

\bibitem[{\citenamefont{Garc\`ia-Aldea and Alvarellos}(2007)}]{alva07}
\bibinfo{author}{\bibfnamefont{D.}~\bibnamefont{Garc\`ia-Aldea}}
  \bibnamefont{and} \bibinfo{author}{\bibfnamefont{J.~E.}
  \bibnamefont{Alvarellos}}, \bibinfo{journal}{J. Chem. Phys.}
  \textbf{\bibinfo{volume}{127}}, \bibinfo{eid}{144109} (\bibinfo{year}{2007}).

\bibitem[{\citenamefont{Perdew and Constantin}(2007)}]{lucianLL}
\bibinfo{author}{\bibfnamefont{J.~P.} \bibnamefont{Perdew}} \bibnamefont{and}
  \bibinfo{author}{\bibfnamefont{L.~A.} \bibnamefont{Constantin}},
  \bibinfo{journal}{Phys. Rev. B} \textbf{\bibinfo{volume}{75}},
  \bibinfo{pages}{155109} (\bibinfo{year}{2007}).

\bibitem[{\citenamefont{Karasiev et~al.}(2009)\citenamefont{Karasiev, Jones,
  Trickey, and Harris}}]{karaPRB09}
\bibinfo{author}{\bibfnamefont{V.~V.} \bibnamefont{Karasiev}},
  \bibinfo{author}{\bibfnamefont{R.~S.} \bibnamefont{Jones}},
  \bibinfo{author}{\bibfnamefont{S.~B.} \bibnamefont{Trickey}},
  \bibnamefont{and} \bibinfo{author}{\bibfnamefont{F.~E.}
  \bibnamefont{Harris}}, \bibinfo{journal}{Phys. Rev. B}
  \textbf{\bibinfo{volume}{80}}, \bibinfo{pages}{245120}
  (\bibinfo{year}{2009}).

\bibitem[{\citenamefont{Karasiev et~al.}(2013)\citenamefont{Karasiev,
  Chakraborty, Shukruto, and Trickey}}]{kara13}
\bibinfo{author}{\bibfnamefont{V.~V.} \bibnamefont{Karasiev}},
  \bibinfo{author}{\bibfnamefont{D.}~\bibnamefont{Chakraborty}},
  \bibinfo{author}{\bibfnamefont{O.~A.} \bibnamefont{Shukruto}},
  \bibnamefont{and} \bibinfo{author}{\bibfnamefont{S.~B.}
  \bibnamefont{Trickey}}, \bibinfo{journal}{Phys. Rev. B}
  \textbf{\bibinfo{volume}{88}}, \bibinfo{pages}{161108}
  (\bibinfo{year}{2013}).

\bibitem[{\citenamefont{Cancio et~al.}(2016)\citenamefont{Cancio, Stewart, and
  Kuna}}]{cancioJCP16}
\bibinfo{author}{\bibfnamefont{A.~C.} \bibnamefont{Cancio}},
  \bibinfo{author}{\bibfnamefont{D.}~\bibnamefont{Stewart}}, \bibnamefont{and}
  \bibinfo{author}{\bibfnamefont{A.}~\bibnamefont{Kuna}}, \bibinfo{journal}{J.
  Chem. Phys.} \textbf{\bibinfo{volume}{144}}, \bibinfo{eid}{084107}
  (\bibinfo{year}{2016}).

\bibitem[{\citenamefont{Perdew et~al.}(1996{\natexlab{b}})\citenamefont{Perdew,
  Burke, and Ernzerhof}}]{pbe}
\bibinfo{author}{\bibfnamefont{J.~P.} \bibnamefont{Perdew}},
  \bibinfo{author}{\bibfnamefont{K.}~\bibnamefont{Burke}}, \bibnamefont{and}
  \bibinfo{author}{\bibfnamefont{M.}~\bibnamefont{Ernzerhof}},
  \bibinfo{journal}{Phys. Rev. Lett.} \textbf{\bibinfo{volume}{77}},
  \bibinfo{pages}{3865} (\bibinfo{year}{1996}{\natexlab{b}}).

\bibitem[{tur()}]{turbomole}
\bibinfo{note}{{TURBOMOLE V6.2, 2009}, a development of {University of
  Karlsruhe} and {Forschungszentrum Karlsruhe GmbH}, 1989-2007, {TURBOMOLE
  GmbH}, since 2007; available from {\tt http://www.turbomole.com}.}

\bibitem[{\citenamefont{Furche et~al.}(2014)\citenamefont{Furche, Ahlrichs,
  H\"attig, Klopper, Sierka, and Weigend}}]{turbo_review}
\bibinfo{author}{\bibfnamefont{F.}~\bibnamefont{Furche}},
  \bibinfo{author}{\bibfnamefont{R.}~\bibnamefont{Ahlrichs}},
  \bibinfo{author}{\bibfnamefont{C.}~\bibnamefont{H\"attig}},
  \bibinfo{author}{\bibfnamefont{W.}~\bibnamefont{Klopper}},
  \bibinfo{author}{\bibfnamefont{M.}~\bibnamefont{Sierka}}, \bibnamefont{and}
  \bibinfo{author}{\bibfnamefont{F.}~\bibnamefont{Weigend}},
  \bibinfo{journal}{Wiley Interdisciplinary Reviews: Computational Molecular
  Science} \textbf{\bibinfo{volume}{4}}, \bibinfo{pages}{91}
  (\bibinfo{year}{2014}).

\bibitem[{\citenamefont{Weigend and Ahlrichs}(2005)}]{def2tzvpp}
\bibinfo{author}{\bibfnamefont{F.}~\bibnamefont{Weigend}} \bibnamefont{and}
  \bibinfo{author}{\bibfnamefont{R.}~\bibnamefont{Ahlrichs}},
  \bibinfo{journal}{Phys. Chem. Chem. Phys.} \textbf{\bibinfo{volume}{7}},
  \bibinfo{pages}{3297} (\bibinfo{year}{2005}).

\bibitem[{\citenamefont{Rappoport and Furche}(2010)}]{furchepol}
\bibinfo{author}{\bibfnamefont{D.}~\bibnamefont{Rappoport}} \bibnamefont{and}
  \bibinfo{author}{\bibfnamefont{F.}~\bibnamefont{Furche}},
  \bibinfo{journal}{J. Chem. Phys.} \textbf{\bibinfo{volume}{133}},
  \bibinfo{pages}{134105} (\bibinfo{year}{2010}).

\bibitem[{\citenamefont{Constantin
  et~al.}(2013{\natexlab{b}})\citenamefont{Constantin, Fabiano, and {Della
  Sala}}}]{bloc}
\bibinfo{author}{\bibfnamefont{L.~A.} \bibnamefont{Constantin}},
  \bibinfo{author}{\bibfnamefont{E.}~\bibnamefont{Fabiano}}, \bibnamefont{and}
  \bibinfo{author}{\bibfnamefont{F.}~\bibnamefont{{Della Sala}}},
  \bibinfo{journal}{J. Chem. Theory. Comput.} \textbf{\bibinfo{volume}{9}},
  \bibinfo{pages}{2256} (\bibinfo{year}{2013}{\natexlab{b}}).

\bibitem[{\citenamefont{del Campo et~al.}(2012)\citenamefont{del Campo,
  G{\'a}zquez, Trickey, and Vela}}]{delCPL12}
\bibinfo{author}{\bibfnamefont{J.~M.} \bibnamefont{del Campo}},
  \bibinfo{author}{\bibfnamefont{J.~L.} \bibnamefont{G{\'a}zquez}},
  \bibinfo{author}{\bibfnamefont{S.}~\bibnamefont{Trickey}}, \bibnamefont{and}
  \bibinfo{author}{\bibfnamefont{A.}~\bibnamefont{Vela}},
  \bibinfo{journal}{Chem. Phys. Lett.} \textbf{\bibinfo{volume}{543}},
  \bibinfo{pages}{179} (\bibinfo{year}{2012}).

\bibitem[{\citenamefont{Zhao and Truhlar}(2005{\natexlab{a}})}]{truhlar05a}
\bibinfo{author}{\bibfnamefont{Y.}~\bibnamefont{Zhao}} \bibnamefont{and}
  \bibinfo{author}{\bibfnamefont{D.~G.} \bibnamefont{Truhlar}},
  \bibinfo{journal}{J. Phys. Chem. A} \textbf{\bibinfo{volume}{109}},
  \bibinfo{pages}{5656} (\bibinfo{year}{2005}{\natexlab{a}}).

\bibitem[{\citenamefont{Zhao and Truhlar}(2005{\natexlab{b}})}]{truhlar05nb}
\bibinfo{author}{\bibfnamefont{Y.}~\bibnamefont{Zhao}} \bibnamefont{and}
  \bibinfo{author}{\bibfnamefont{D.~G.} \bibnamefont{Truhlar}},
  \bibinfo{journal}{J. Chem. Theory and Comput.} \textbf{\bibinfo{volume}{1}},
  \bibinfo{pages}{415} (\bibinfo{year}{2005}{\natexlab{b}}).

\bibitem[{\citenamefont{Laricchia et~al.}(2013)\citenamefont{Laricchia,
  Fabiano, and {Della Sala}}}]{fde_ct}
\bibinfo{author}{\bibfnamefont{S.}~\bibnamefont{Laricchia}},
  \bibinfo{author}{\bibfnamefont{E.}~\bibnamefont{Fabiano}}, \bibnamefont{and}
  \bibinfo{author}{\bibfnamefont{F.}~\bibnamefont{{Della Sala}}},
  \bibinfo{journal}{J. Chem. Phys.} \textbf{\bibinfo{volume}{138}},
  \bibinfo{eid}{124112} (\bibinfo{year}{2013}).

\bibitem[{\citenamefont{Fabiano
  et~al.}(2014{\natexlab{b}})\citenamefont{Fabiano, Constantin, and {Della
  Sala}}}]{dihydrogen}
\bibinfo{author}{\bibfnamefont{E.}~\bibnamefont{Fabiano}},
  \bibinfo{author}{\bibfnamefont{L.~A.} \bibnamefont{Constantin}},
  \bibnamefont{and} \bibinfo{author}{\bibfnamefont{F.}~\bibnamefont{{Della
  Sala}}}, \bibinfo{journal}{J. Chem. Theory. Comput.}
  \textbf{\bibinfo{volume}{10}}, \bibinfo{pages}{3151}
  (\bibinfo{year}{2014}{\natexlab{b}}).

\bibitem[{\citenamefont{Laricchia et~al.}(2012)\citenamefont{Laricchia,
  Fabiano, and {Della Sala}}}]{fde_hyb_ene}
\bibinfo{author}{\bibfnamefont{S.}~\bibnamefont{Laricchia}},
  \bibinfo{author}{\bibfnamefont{E.}~\bibnamefont{Fabiano}}, \bibnamefont{and}
  \bibinfo{author}{\bibfnamefont{F.}~\bibnamefont{{Della Sala}}},
  \bibinfo{journal}{J. Chem. Phys.} \textbf{\bibinfo{volume}{137}},
  \bibinfo{eid}{014102} (\bibinfo{year}{2012}).

\bibitem[{sup()}]{suppmat}
\bibinfo{note}{See supplementary material at
  http://aip.scitation.org/doi/suppl/10.1063/1.4975092.}

\bibitem[{\citenamefont{Tran and Wesoloski}(2002)}]{TW02}
\bibinfo{author}{\bibfnamefont{F.}~\bibnamefont{Tran}} \bibnamefont{and}
  \bibinfo{author}{\bibfnamefont{T.~A.} \bibnamefont{Wesoloski}},
  \bibinfo{journal}{Int. J. Quant. Chem.} \textbf{\bibinfo{volume}{89}},
  \bibinfo{pages}{441} (\bibinfo{year}{2002}).

\bibitem[{\citenamefont{Lembarki and Chermette}(1994)}]{lc94}
\bibinfo{author}{\bibfnamefont{A.}~\bibnamefont{Lembarki}} \bibnamefont{and}
  \bibinfo{author}{\bibfnamefont{H.}~\bibnamefont{Chermette}},
  \bibinfo{journal}{Phys. Rev. A} \textbf{\bibinfo{volume}{50}},
  \bibinfo{pages}{5328} (\bibinfo{year}{1994}).

\bibitem[{\citenamefont{Becke and Edgecombe}(1990)}]{elf1}
\bibinfo{author}{\bibfnamefont{A.~D.} \bibnamefont{Becke}} \bibnamefont{and}
  \bibinfo{author}{\bibfnamefont{K.~E.} \bibnamefont{Edgecombe}},
  \bibinfo{journal}{J. Chem. Phys.} \textbf{\bibinfo{volume}{92}},
  \bibinfo{pages}{5397} (\bibinfo{year}{1990}).

\bibitem[{\citenamefont{Silvi and Savin}(1994)}]{elf2}
\bibinfo{author}{\bibfnamefont{B.}~\bibnamefont{Silvi}} \bibnamefont{and}
  \bibinfo{author}{\bibfnamefont{A.}~\bibnamefont{Savin}},
  \bibinfo{journal}{Nature} \textbf{\bibinfo{volume}{371}},
  \bibinfo{pages}{683} (\bibinfo{year}{1994}).

\bibitem[{\citenamefont{Pittalis et~al.}(2015)\citenamefont{Pittalis, Troiani,
  Rozzi, and Vignale}}]{pittalis15}
\bibinfo{author}{\bibfnamefont{S.}~\bibnamefont{Pittalis}},
  \bibinfo{author}{\bibfnamefont{F.}~\bibnamefont{Troiani}},
  \bibinfo{author}{\bibfnamefont{C.~A.} \bibnamefont{Rozzi}}, \bibnamefont{and}
  \bibinfo{author}{\bibfnamefont{G.}~\bibnamefont{Vignale}},
  \bibinfo{journal}{Phys. Rev. B} \textbf{\bibinfo{volume}{91}},
  \bibinfo{pages}{075109} (\bibinfo{year}{2015}).

\end{thebibliography}

\end{document}